\documentclass[twocolumn]{aastex631}
\usepackage{showyourwork}
\usepackage{rsfso}
\usepackage{rotating}

\newcommand{\angstrom}{\mbox{\normalfont\AA}}
\newcommand{\nflares}{26,355}
\newcommand{\nflarestars}{3,160}
\newcommand{\nstars}{6,824}
\newcommand{\nprot}{1,847}

\begin{document}

\title{Evolution of Flare Activity in GKM Stars Younger than 300 Myr over Five Years of TESS Observations}

\shorttitle{Evolution of young stellar flare rates}
\shortauthors{Feinstein et al.}

\author[0000-0002-9464-8101]{Adina~D.~Feinstein}
\altaffiliation{NHFP Sagan Fellow}
\affiliation{Laboratory for Atmospheric and Space Physics, University of Colorado Boulder, UCB 600, Boulder, CO 80309}

\author[0000-0002-0726-6480]{Darryl~Z.~Seligman}
\altaffiliation{NSF Astronomy and Astrophysics Postdoctoral Fellow}
\affiliation{Department of Astronomy and Carl Sagan Institute, Cornell University, 122 Sciences Drive, Ithaca, NY 14853}

\author[0000-0002-1002-3674]{Kevin~France}
\affiliation{Laboratory for Atmospheric and Space Physics, University of Colorado Boulder, UCB 600, Boulder, CO 80309}
\affiliation{Department of Astrophysical and Planetary Sciences, University of Colorado, UCB 389, Boulder, CO 80309, USA}
\affiliation{Center for Astrophysics and Space Astronomy, University of Colorado, 389 UCB, Boulder, CO 80309, USA}

\author[0000-0002-2592-9612]{Jonathan Gagné}
\affiliation{Planétarium Rio Tinto Alcan, Escape pour la Vie, 4801 av. Pierre-de Coubertin, Montréal, Québec, Canada}
\affiliation{Institute for Research on Exoplanets, Université de Montréal, Départment de Physique, C.P. 6128 Succ. Centre-ville, Monetréal, QC H3C 3J7, Canada}

\author[0000-0001-7458-1176]{Adam~Kowalski}
\affiliation{Department of Astrophysical and Planetary Sciences, University of Colorado, UCB 389, Boulder, CO 80309, USA}
\affiliation{National Solar Observatory, University of Colorado Boulder, 3665 Discovery Drive, Boulder CO 80303, USA}

\begin{abstract}

Stellar flares are short-duration ($<$\,hours) bursts of radiation associated with surface magnetic reconnection events.
Stellar magnetic activity generally decreases as a
function of both age and Rossby number, $R_0$, a measure of the relative importance of the convective and
rotational dynamos.  Young stars ($<300$ Myr) have typically been
overlooked in population-level flare studies due to challenges with
flare-detection methods. Here, we select a sample of stars
that are members of 26 nearby  moving groups, clusters, or associations with ages $<$300~Myr  that have been observed
by the Transiting Exoplanet Survey Satellite (TESS) at
2-minute cadence.  We identified \nflares\ flares originating from \nflarestars\
stars and robustly measure the rotation periods of \nprot~stars. We measure and find the flare frequency
distribution (FFD) slope, $\alpha$,
saturates for all spectral types at $\alpha \sim -0.5$ and is constant over 300~Myr. Additionally, we find that
flare rates for stars $t_\textrm{age} = 50 - 250$~Myr are
saturated below $R_0 < 0.14$, which is consistent with other indicators of magnetic activity. We find evidence
of annual flare rate variability in eleven stars, potentially correlated with long term stellar activity cycles.
Additionally, we cross match our entire sample  with \textit{GALEX} and find no
correlation between flare rate and Far- and Near-Ultraviolet flux.
Finally, we find the flare rates of planet hosting stars are relatively lower than comparable, larger samples
of stars, which may have ramifications for the atmospheric
evolution of short-period exoplanets.

\end{abstract}

\section{Introduction} \label{sec:intro}

Stellar flares are the radiation component of magnetic reconnection events \citep{benz10}. Such events are readily seen on the Sun \citep{carrington1859, lu91, fletcher11}, particularly during the maximum in the solar cycle \citep{webb94}. Solar flares can be used as proxies for magnetic activity occurring on other stars \citep{feigelson99, berdyugina05, kowalski10, feinstein22_criticality}. Additionally, these short-duration flaring events can have significant ramifications on the evolution of short-period extrasolar planets \citep{france16, guenther19_flares, chen21}. While stellar flares are typically not spatially resolvable, they do lend themselves to characterization via spectroscopic and photometric signatures. Spectroscopic characterization of stellar flares inform our understanding of non-thermal processes affiliated with such events such as coronal mass ejections \citep{argiroffi19, vida19}, proton beams \citep{orrall76, woodgate92}, and accelerated electrons \citep{osten05, smith05}.

Photometric observations of stars are more readily available now with exoplanet transit discovery missions, and allow us to statistically characterize flare rates and energies at optical/near-infrared wavelengths. Observations of M dwarfs with the Sloan Digital Sky Survey revealed a correlation between the flaring fraction of stars with height above the galactic plane, a proxy for age \citep{kowalski09, hilton10}. More recently, NASA's \textit{Kepler}, \textit{K2}, and the Transiting Exoplanet Survey Satellite (TESS) missions have provided a wealth of stellar variability data in addition to their primary objective of detecting transiting exoplanets. Flares can be identified within time-series photometry by a sharp rise and subsequent exponential decay in flux,  with the latter corresponding to the cooling rate \citep{kowalski13}. \textit{Kepler} \citep{Borucki10} provided long-baseline high-cadence observations used to identify stellar flares. There has been extensive studies of flares in \textit{Kepler} data, from the statistics of superflares on solar-type stars \citep[e.g.][]{notsu13, shibayama13, maehara15, okamoto21} to low-mass stars \citep[e.g.][]{Hawley14, silverberg16}. \cite{davenport19} found that flare activity decreased with increasing rotation period for 347 GKM stars. However, the flare frequency distribution (FFD) slope did not vary significantly as a function of age. As a caveat, the ages of the stars in this study were determined based on their rotation periods, relying on the assumption that gyrochronology alone accurately ages stars.

\textit{K2} provided 70-day baseline observations for a handful of young stars in groups such as Upper Scorpius,  Pleiades, Hyades, and Praespe clusters. \cite{ilin19, ilin21} analyzed flares from K and M stars in these clusters and found that the overall flare activity decreased as a function of age. Moreover, this relationship was steeper for more massive stars in the sample. \cite{paudel18} surveyed 10 M6 - L0 dwarfs observed with \textit{K2} and found the L0 dwarfs had significantly shallower FFDs than the M dwarfs. They found that, on average, young targets (defined by the tangential velocity of the star) exhibited more flares overall. More recently, TESS \citep{ricker15} has provided near all-sky photometric observations at 30-minute cadence or less. This observing strategy has allowed for more detailed studies of young stellar flares from nearby, disperse young moving groups and associations. These  data permit detailed studies of individual stars, for example characterizing eight superflares ($E_f > 10^{34}$~erg) on the young star AB Doradus over $\sim 60$ days of continuous observations \citep{schmitt19}, as well as statistical studies of flares across a range of spectral types and ages \citep{doyle20, feinstein22_criticality, pietras22, yang23}.

The 11-year solar activity cycle represents a change in the magnetic activity of the Sun, and manifests itself in a variety of observables including increases in the total number of sunspots \citep{clette14, kilcik14}, flares and coronal mass ejections \citep{crossby93, webb94, lin23}, and an increase in the total solar irradiance \citep{lean1987}. Insights into the long-term activity cycles on other stars have been limited to photometric and spectroscopic monitoring \citep{saar99}. \cite{lehtinen16} collected 16 to 27 years of B- and V-band photometry for 21 young active solar-type stars. They found evidence of long-term stellar activity cycles in 18 targets for which$P_\textrm{rot}/P_\textrm{cycle} \propto R_0^{-1}$, where $P_\textrm{rot}$ is the rotation period, $P_\textrm{cycle}$ is the period of the stellar activity cycle, and $R_0$ is the Rossby number. Additionally, 50 years of spectroscopic observations of HD~166620 from the Mount Wilson Observatory and Keck revealed both a $\sim 16$~year periodicity in emission from the core of the \ion{Ca}{2} H and K lines \citep{olah16} and evidence that this star entered a grand minima \citep{baum22}. A more complete review on the state of stellar activity cycles can be found in \cite{jeffers23} and \cite{isik23}.

In addition to photometric and spectroscopic observations, solar flares trace the length of the solar activity cycle. The high-cadence observations from \textit{Kepler} (4-year baseline) and TESS (currently 5-year baseline) provide  sufficient data for searches of stellar activity cycles from the variations in stellar flare rates. \cite{davenport20} demonstrated that the flare rate and FFDs of GJ~1234 has not changed appreciably over 10~years of observations with both \textit{Kepler} and TESS. On the other hand, \cite{scoggins19} found that the M3V star star KIC 8507979 showed a clear decline in flare rate and change in FFD over 4-years of \textit{Kepler} observations. While it is not expected to find flare rate and distribution variations in all stars given detection limitations, KIC 8507979 demonstrates the ability to study long-term flare variability as a potential tracer for stellar activity cycles.

The paper is presented as follows. In Section~\ref{sec:methods}, we describe our sample,
and stellar flare and rotation period identification methods. In Section~\ref{sec:results},
we present our flare-frequency distribution (FFD) fits as a function of stellar age, $T_\textrm{eff}$,
and $R_0$. In Section~\ref{sec:cycles}, we present evidence of flare rate changes over
the 5-year TESS baseline in eleven young stars, likely correlated with long term stellar
activity cycles. In Section~\ref{sec:discuss}, we search for correlations in flare rates
with FUV and NUV observations from \textit{GALEX} and place the flare rates of young planet
hosting stars in the context of our broader sample.  We conclude in Section~\ref{sec:conclusions}.
We provide additional figures and tables in Section~\ref{appendix:supp_ffds}.

This paper was written using the \textcolor{red}{\textit{showyourwork!}} open-source
software package. The objective of \textcolor{red}{\textit{showyourwork!}} is to improve reproducibility
and transparency of scientific research by compiling the manuscript and figures
simultaneously. All of the data in this work is hosted on GitHub\footnote{The GitHub is
hyperlinked in the \textcolor{red}{\textit{showyourwork!}} stamp at the top of the first page of this manuscript,
 and can also be found here: \url{https://github.com/afeinstein20/young-stellar-flares}.} and
 Zenodo.\footnote{Link to be uploaded upon acceptance.} At the end of every caption figure in this manuscript,
 there is a GitHub icon (\faGithub), which links to the Python script used to create that figure.

\section{TESS Light Curve Characterization}\label{sec:methods}

Here, we provide an overview of the methodology used in this paper. Specifically, we describe
the sample selection in Section~\ref{subsec2:sample}, TESS light curve analysis in
Section~\ref{subsec2:TESS_lightcurve}, flare identification in Section~\ref{subsec2:Flareidentification},
flare fitting parameters in Section~\ref{subsec2:model}, flare quality checks in
Section~\ref{subsec2:qualitychecks}, and  stellar rotation period measurements in Section~\ref{subsec:prot}.

\subsection{Sample Selection}\label{subsec2:sample}

A primary goal of this paper is to measure the relationship between the flare rates and
ages of stars. We are particularly interested in this dependency for young stars with ages
$4 \leq t_\textrm{age} \leq 300$\,Myr. To this end, we used the MOCA Data Base (Gagné et al.
in prep.)\footnote{\url{https://mocadb.ca/}} to identify 26~ nearby, aged, young moving groups,
associations, and open clusters from which we created our sample of stars. The final targets
were required to be: (i) confirmed members, (ii) high-likelihood candidate members, or (iii)
candidate members.  Membership to these groups has been primarily determined using kinematic
information from \textit{Gaia} \citep{gaia16, gaia18}. The membership status is determined by
the probability that BANYAN-$\Sigma$ assigns based on how well the kinematics of the target
matches with the kinematics of the group \citep{gagne18}. These rather stringent cuts resulted
in a catalog of 30,889 stars across 26 associations. We summarize the sample and ages for each
association (and therefore star) in Table~\ref{tab:sample}.

\begin{deluxetable}{l r r r r r}[!ht]
\tabletypesize{\footnotesize}
\tablecaption{Adopted Ages of each Young Stellar Population and Number of
Stars per Group, $N_\textrm{stars}$ \label{tab:sample}}
\tablehead{
\colhead{Population} & \colhead{Age [Myr]} & \colhead{N$_\textrm{stars}$} &
\colhead{Ref.}
}
\startdata
AB Doradus  &  133$_{-20}^{+15}$  &  88  & 1 \\
Blanco 1 & 137.1$^{+7.0}_{-33}$ & 428 & 2\\
Carina  &  45$\pm 9$  & 94  & 3\\
Carina-Musca & 32 & 35 & 4\\
Chamaeleon  &  5  &  424  & 5 \\
Columba  &  42  &  126  & 3\\
Greater Taurus Subgroup 5 & 8.5 & 56 & 4\\
Greater Taurus Subgroup 8 & 4.5 & 122 & 4\\
Lower Centaurus Crux & 15 & 761 & 6\\
MELANGE-1 & 250$_{-70}^{+50}$ & 19 & 7\\
Octans  &  35$\pm 5$ &   64  & 8 \\
Pisces Eridanis  &  120  &   219  & 9 \\
Pleiades & 127.4$^{+6.3}_{-10}$ & 1421 & 2\\
$\alpha$ Persei  &  79$_{-2.3}^{+1.5}$  &  625 & 2 \\
IC 2602 system &  52.5$_{-3.7}^{+2.2}$  & 160 & 2 \\
NGC 2451A & 48.5 & 59 & 4\\
Oh 59 & 162.2 & 62 & 10\\
Platais 9 & 50 & 124 & 11\\
RSG2 & 126 & 145 & 12\\
Theia 301  &  195  &  437 & 10\\
Theia 95  &  30.2  &  230 & 10\\
TW Hydrae  &  10  &  24  & 3\\
Upper Centaurus Lupus & 16 $\pm$ 2 & 696 & 6\\
Upper Scorpius  &  10  &  106  &  6\\
Vela-CG4  &  33.7  &  299  & 4\\
\hline
Total & & \nstars\\
\enddata
\tablecomments{Age references: (1) \citet{gagne18_abdmg}; (2) \citet{galindo22}; (3) \citet{Bell15}; (4) \citet{kerr21}; (5) \citet{luhman07}; (6) \citet{pecaut16}; (7)
\citet{tofflemire21}; (8) \citet{murphy15}; (9) \citet{curtis19}; (10) \citet{kounkel20}; (11) \citet{tarricq21}; (12) \citet{roser16}}
\end{deluxetable}

\subsection{TESS Light Curves}\label{subsec2:TESS_lightcurve}

We cross-matched our MOCA Database sample with the TESS Input Catalog (TIC) based
on their Gaia DR2 RA and Dec; we required that the distance between the target and
the nearest TIC target was within $<1"$. We down-selected our sample to stars that
have been observed with TESS at 2-minute cadence. This ensures that we are able to
temporally resolve and accurately measure the properties of flares for each of these
stars \citep{howard22}.

\begin{figure}[ht!]
    \script{sample.py}
    \begin{centering}
        \includegraphics[width=\linewidth]{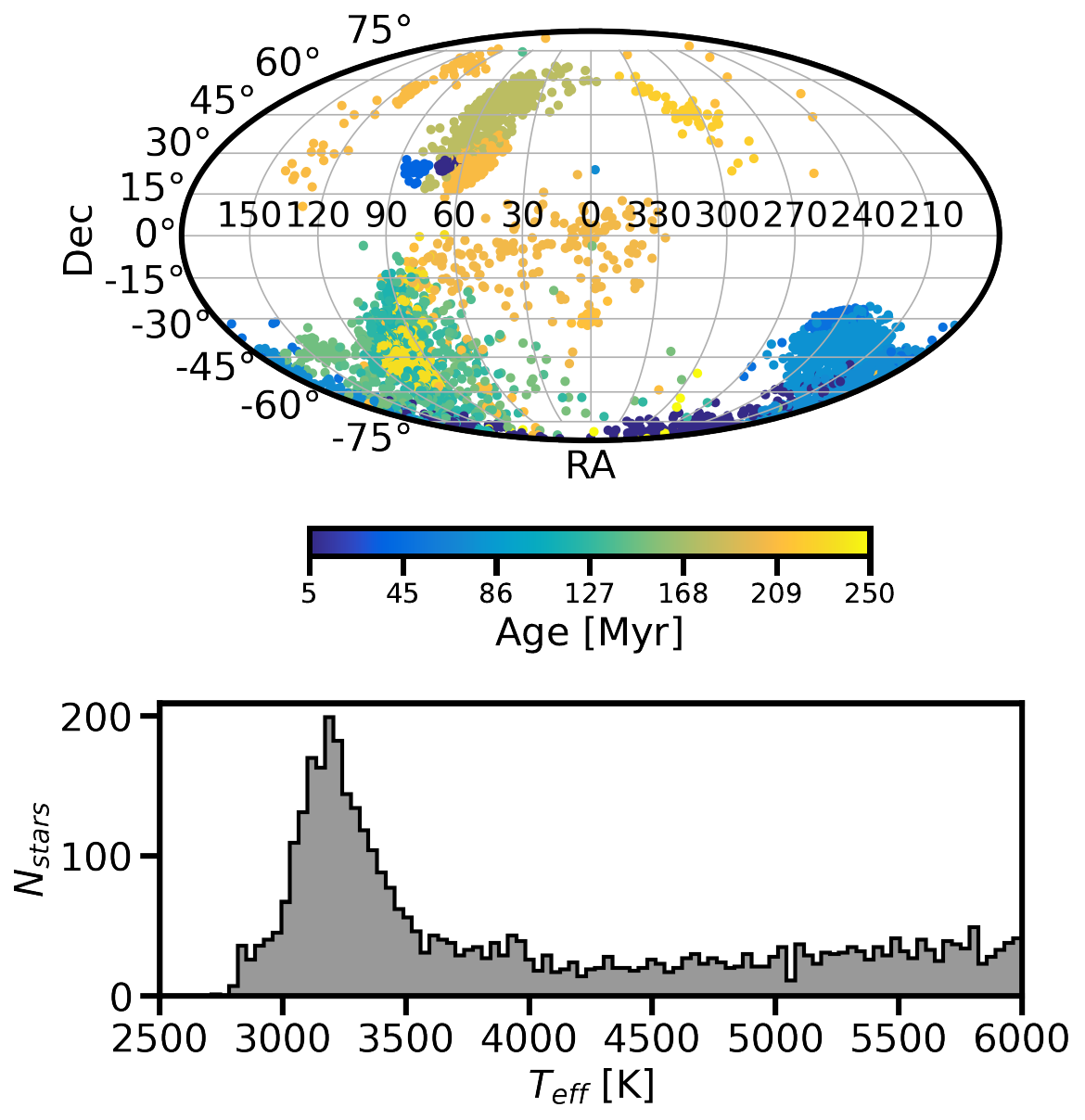}
        \caption{
            Top: Distribution of the selected sample across the sky and colored by the adopted age of the association (see Table~\ref{tab:sample}). The all-sky coverage by TESS
            has unlocked new populations of stars to observe. We take advantage of this observing strategy to measure flare rates across 26 different nearby young moving groups,
            clusters, and associations. Bottom: Distribution of adopted effective temperatures, $T_\textrm{eff}$ [K] for stars in our sample. We include all stars with $T_\textrm{eff} \leq 6000$\,K.
        }
        \label{fig:sample}
    \end{centering}
\end{figure}

This process provided a final sample of \nstars\ unique targets. These targets
have each been observed at a 2-minute cadence between TESS Sector 1 and Sector 67;
Sector 67 was the latest available sector at the time that this analysis was performed.
In Figure ~\ref{fig:sample}, we show the distribution of the on sky positions, ages
and effective temperatures, $T_\textrm{eff}$, of the final sample. Because many of
the groups are located at or near the ecliptic poles, many of our targets were observed
over multiple TESS sectors. We use $T_\textrm{eff}$ as our stellar characterization
metric over the standard $M_\star$ because $\sim 600$ stars in our sample lack mass
information in the TIC. The entirety of the available data yielded 17,964 light curves
processed by the Science Processing Operations Center (SPOC) pipeline \citep{jenkins16},
which can be accessed on MAST under DOI  \href{http://dx.doi.org/10.17909/t9-nmc8-f686}{10.17909/t9-nmc8-f686}.
Our sample has an average of $\sim 3$ light curves per target (although with significant
spread across targets). We downloaded all light curves using \texttt{lightkurve}\footnote{\url{https://doi.org/10.5281/zenodo.4654522}}.
For our analysis, we used the SPOC-processed SAP\_FLUX.

\subsection{Flare Identification}\label{subsec2:Flareidentification}

Once we downloaded the light curves for each target, we performed the following
procedure to identify stellar flares. We implemented the machine learning
flare-identification methods presented and described in \cite{feinstein20}.
This method relies on the similar time-dependent morphologies of all flare events.
These flare-profiles can generally be described as a sharp rise followed by an
exponential decay in the white light curve. This identification-technique implements
the convolutional neural network (CNN) \texttt{stella} \citep{feinstein20},
although other architectures have been explored for flare identification
\citep[e.g.][]{vida18}. The CNN  was trained on a by-eye validated catalogue of flares
from TESS Sectors 1 and 2 with 2-minute data \citep{guenther19_flares}.

There are several benefits to using the CNN for stellar flare identification. Primarily,
the CNN is insensitive to the stellar baseline flux because it is trained to search
\textit{only} based on flare morphology. It is therefore relatively insensitive to
the absolute flux levels, so long as the inherent noise does not overwhelm the signal
itself. An additional benefit is that rotational modulation peaks --- which are
themselves driven by stellar heterogeneities --- are not accidentally identified as
flares. This holds true for stars with rotation periods, $P_\textrm{rot} > 1$~day.
This is especially advantageous for our sample of young stars, which readily exhibit
rotational modulation in their light curves.

Based on these advantages, the final compiled sample of flares is unbiased towards
low-amplitude/low-energy flares. It is important to note that these low-energy events
are typically not identified in traditional sigma-outlier identification
methods \citep[e.g.][]{chang15, vasilyev22}. While
significant developments have been achieved to better model and detrend stellar
variability \citep[e.g.][]{bicz22}, these methods can be computationally intensive
The \texttt{stella} CNN models calculate the probability that a data point in a light
curve is associated with a flaring event. Specifically, it takes the light curve (time,
flux, flux error) as an input and returns an array with values of [0,1], which are
treated as the probability a data point is (1) or is not (0) part of a flare. We ran
every light curve through 10 independent \texttt{stella} models and averaged the
outputs to ensure that our statistics were accurate. We note that in this processing,
the CNN ignores 200-minutes before and after any gaps in the data. Therefore, any
flares which occur during these times are not identified. From by-eye
vetting of $\sim 300$ light curves, we found 2~flares within the first 200~minutes of
the orbital gaps. Therefore, we estimate $\leq 120$~flares are missed in our catalog.

The \texttt{stella} code groups individual points with the predictions per data point
when identifying a single flare event. We modified this stage of identification slightly
from the original flare-identification method. Specifically, we identified all data
points with a probability of being associated with a flare of $P > 0.75$. Any data
points that were within 4 cadences of each other were considered to be a single flare
event. We did not consider any potential flares that had three or fewer points with
$P > 0.75$. This method rejects single-point outliers which can be assigned  high
probabilities of being  flares. We assigned the probability of the whole flare event
as the probability of the peak data point.

\begin{figure}[ht!]
    \script{flare_distribution.py}
    \begin{centering}
        \includegraphics[width=0.9\linewidth]{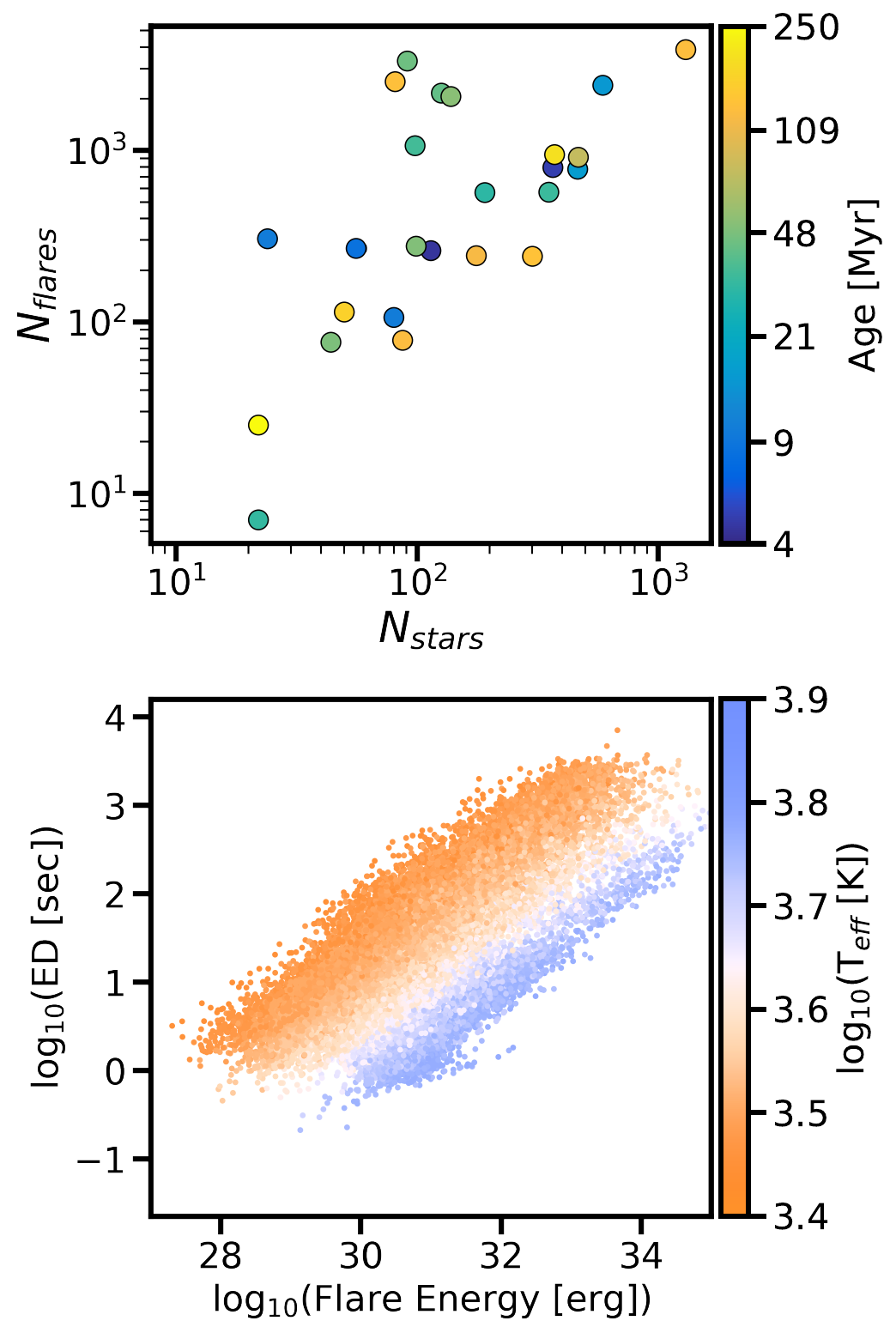}
        \caption{
            High level summary of the demographics of flares in our sample. Top: The number of flares identified compared to the number of stars in each nearby young moving group, cluster,
            or association. A linear relationship is expected. Bottom: The distribution of measured TESS energies
            and equivalent durations of flares in our sample, colored by the probability of the flare as identified with \texttt{stella}.
        }
        \label{fig:flare_distribution}
    \end{centering}
\end{figure}

\subsection{Modeling Flare Properties}\label{subsec2:model}

Flares are well-described in light curves as a sudden increase in flux followed
by an exponential decay. We used the analytic flare model, Llamaradas Estelares
which was presented in \cite{tovar22}\footnote{\url{https://github.com/lupitatovar/Llamaradas-Estelares}},
to fit and extract parameters of flares in all of the light curves of our sample.
This model builds upon the model presented in \cite{davenport14}. Specifically,
it includes the convolution of a Gaussian with a double exponential model in the
flare profile.

The analytic model robustly  accounts for physically-motivated flare features.
Specifically, it can incorporate the flare (i) amplitude, (ii) heating timescale,
(iii) rapid cooling phase timescale, and (iv) slow cooling phase timescale. We
implemented a nonlinear least squares optimization to fit the flare peak time
($t_\textrm{peak}$), full width at half maximum (FWHM), and the amplitude ($A$)
of each flare in the sample. We note that there are several other
physically-motivated flare models which could be used, such as those presented in
\cite{gryciuk17, pietras22, yang23_lp}; these models include using two profiles to
fit the impulsive and late phases of the flare. We combine the model with a
second-order polynomial fit to a 1.2-hour baseline before and after the flare
during only the fitting stage. This was implemented in order to account for any
slope due to rotational modulation, and therefore was particularly relevant for the
rapid rotators ($P_\textrm{rot} < 2$~days).

We calculated the equivalent duration, ED, of the flare by integrating the
quiescent-normalized flare flux with respect to time. We calculate the flare
energy, $E_\textrm{flare}$, using,

\begin{equation}\label{eq:Eflare}
    E_\textrm{flare} = L_\star \, A \, \textrm{ED} \, s \, .
\end{equation}
In Equation \ref{eq:Eflare}, $L_\star$ is the luminosity of the star and $s$ is a
scaling factor defined as $s = B_\lambda(T_\textrm{eff}) / B_\lambda(T_\textrm{flare})$,
where $B_\lambda(T)$ is the Planck function. We assume that the flare temperature is
9000~K \citep{hawley92, hawley95}, although recent NUV flare observations suggest
this may be an underestimation \citep{kowalski19, brasseur23, berger23}.

\subsubsection{Comparison of Flare Models}

\begin{figure}[ht!]
    \script{model_fit.py}
    \begin{centering}
        \includegraphics[width=\linewidth]{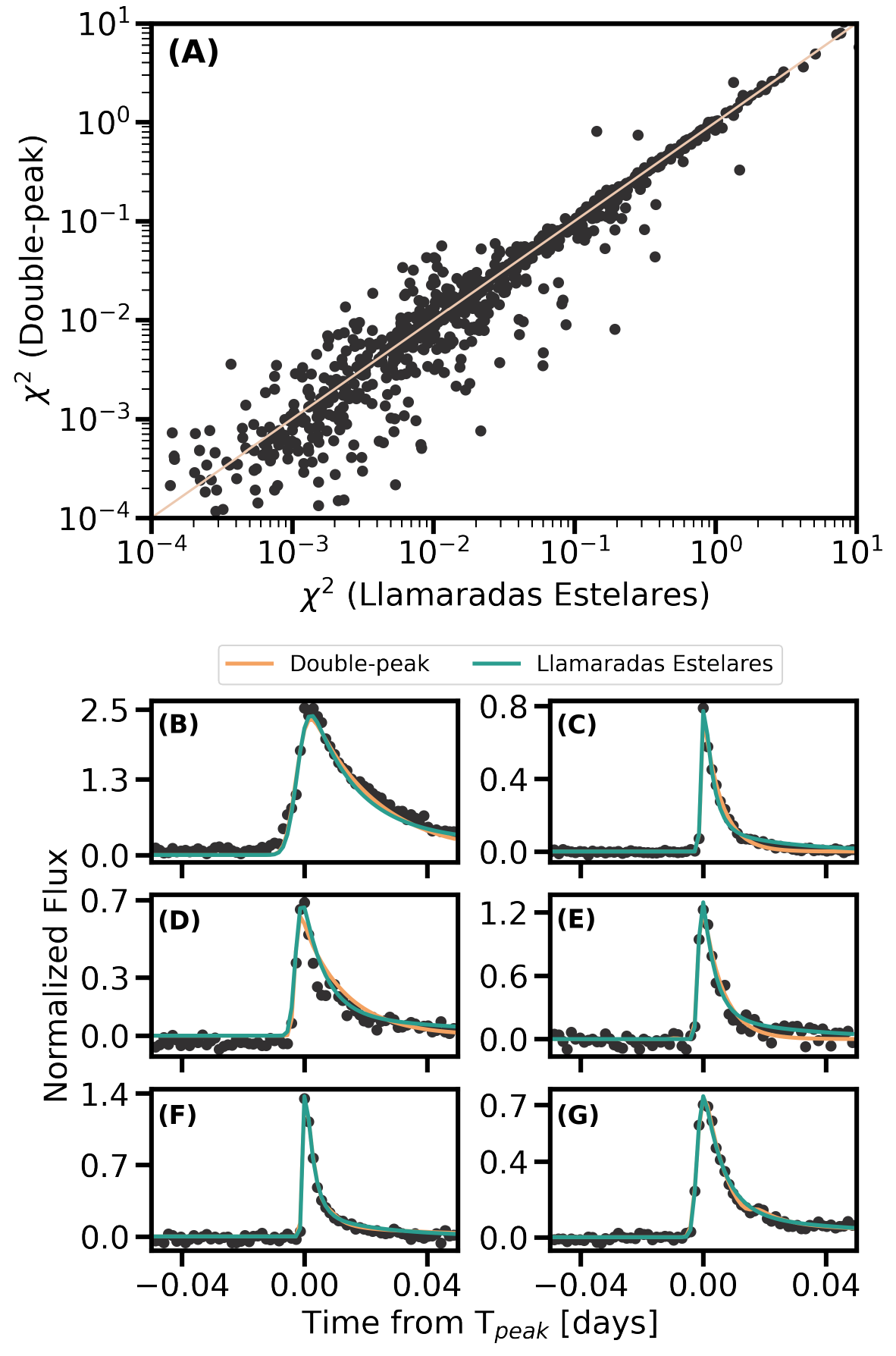}
        \caption{
            Comparison fits between the Llamaradas Estelares
            single-peak model from \cite{tovar22} and the double-peak flare model
            from \cite{pietras22} for flares with $A \geq 0.4$. Top (A): Comparison
            in the $\chi^2$ between the best-fit model for 900 flares. A one-to-one
            line is plotted in peach to guide the eye. Across this sample of
            high-amplitude flares, there is good agreement between the fits for
            these two distinct models. Bottom: Comparison in best-fit models for
            a subset of flares. The best-fit Llamaradas Estelares model is plotted
            in blue; the best-fit double-peak model is plotted in orange. We
            highlight examples where the double-peak flare model has the better
            fit (B, C), the Llamaradas Estelares flare model has the better fit
            (D, E), and where the models perform equally as well (F, G). We note
            there is no correlation between flare amplitude and preferred model.
        }
        \label{fig:model_fit}
    \end{centering}
\end{figure}

\cite{tovar22} demonstrated that the flare model works well for
flares with amplitudes $\geq 2$ (i.e. double the baseline flux). However, other
works such as \cite{pietras22} have demonstrated that a double-flare profile better
fits high energy flares. To quantify how well the Llamaradas Estelares model works,
we refit and calculate the $\chi^2$ for a subset of high amplitude flares ($A \geq 0.4$)
using Equation 3 in \cite{pietras22}, a double-peak flare model, which is:

\begin{equation}\label{eqn:flare}
\begin{split}
    f(\tau) = \int_0^\tau\bigg( A_1\, \exp\bigg[\frac{-(t-B_1)^2}{C_1^2}\bigg] \,\exp\bigg[\frac{(t-\tau)}{D_1}\bigg]\\+A_2 \,\exp\bigg[\frac{-(t-B_2)^2}{C_2^2}\bigg] \,\exp\bigg[\frac{(t-\tau)}{D_2}\bigg]\bigg) \textrm{d}t
\end{split}
\end{equation}

where $A$ is is related to amplitude of the flare, $B$ relates
to the total energy released during the flare, $C$ relates to the timescale of
the flare, $D$ is related to the decay timescale, $\tau$ is the time of a given
cadence, and $t$ is the time to integrate over. The subscripts 1 and 2 indicate
which flare variables are associated with the first and second flare profile. We
refit 900~high-amplitude flares following Equation~\ref{eqn:flare}. We plot the
results of this refitting in Figure~\ref{fig:model_fit}.

Broadly, we find the flare model from \cite{tovar22} and \cite{pietras22}
to fit the majority of the high-amplitude flares equally as well (Figure~\ref{fig:model_fit}\,A).
We highlight several examples where one model fits better than the other in Figure~\ref{fig:model_fit}\,B-E,
and cases where both profiles fit the data equally as well Figure~\ref{fig:model_fit}\,F-G.
Flares which prefer the Llamaradas Estelares model tend to have a more accurately fit flare
amplitude. Flares which prefer the double-peak model tend to have an extended decay timescale,
which is better fit by the addition of a secondary flare. Flares which are fit equally well by
both models tend to have a better fit to the amplitude in the Llamaradas Estelares model, but a
better fit to the decay timescale in the double-peak flare model.

\subsection{Flare Quality Checks}\label{subsec2:qualitychecks}

The \texttt{stella} CNNs were trained on data from TESS Sectors 1 and 2. However, the
noise properties are variable across sectors in TESS data. The CNNs are therefore unable
to accurately account for and capture  this variation when operating on different sectors.
Moreover, the original CNNs were only trained on a sample of 1,228 stars. This training
sample does not necessarily encapsulate a sufficient distribution  of variable stars types,
such as eclipsing binaries, RR Lyraes, and fast rotators with $P_\textrm{rot} < 1$~day
\citep{lawson19}, which results in the CNNs not being able to properly distinguish these
temporal features from flares. We therefore apply additional quality checks to ensure our
flare sample has little to no contamination from other sources.

Specifically, we remove flares from our sample which did not satisfy  one or more of the following criteria:
\begin{enumerate}
    \item  The amplitude of the flare must be $> 0.01$, the same limits set by \cite{feinstein20}.
    \item The flare amplitude  must be at least twice the standard deviation of the light curve
    30 minutes before and 45 minutes after the flare. This ensures that the feature is not a sharp noise artifact.
    \item  The fitted flare model parameters must be physically motivated: FWHM $> 0$; $A > 0$; ED $> 0$.
    \item The flare parameters must be $\sigma_A < 0.5$ and $\sigma_{t_\textrm{peak}} < 0.01$.
\end{enumerate}

We find that flares are often simply mischaracterized noise when the errors on the first
and fourth criteria are larger than the cut-offs. These quality checks highlight the
need to continuously update machine learning models, especially when looking at data
with varying instrumental systematics.

As a final check, we performed an exhaustive by-eye verification of flares from light
curves for stars with flare rates, $\mathcal{R} > 1$~day$^{-1}$. These stars generally
tend to have $T_\textrm{eff} > 5000$~K and TESS magnitudes $< 8$. Therefore, their
light curves are dominated by sharp noise which \texttt{stella} often mischaracterizes
as flares. As a final cut, the sample only includes events that have  a probability
$P \geq 98\%$ of being a true flare. After performing these additional checks, we
obtain a robust final flare sample of \nflares\ flares originating from \nflarestars\
stars (Figure~\ref{fig:flare_distribution}).

\subsection{Measuring Rotation Periods}\label{subsec:prot}

In addition to understanding flare statistics across young stars, we measure the
rotation periods, $P_\textrm{rot}$ of stars in our sample. We then search for correlations
between $P_\textrm{rot}$ and Rossby number, $R_0$. \citet{seligman22} demonstrated that
stars with low Rossby numbers $R_0 < 0.13$ exhibit shallower flare frequency distribution
slopes. These shallower slopes are caused by  an excess of  high energy flares compared
to lower energy flares.

In order to perform this, we describe how we measure stellar rotation periods from
the TESS light curves. To this end we used \texttt{michael}\footnote{\url{https://github.com/ojhall94/michael}},
an open-source Python package that robustly measures $P_\textrm{rot}$ using a combination
traditional Lomb-Scargle periodograms and wavelet transformations (Hall et al. submitted).
\texttt{michael} measures $P_\textrm{rot}$ using the \texttt{eleanor} package, which
extracts light curves from the TESS Full-Frame Images \citep[FFIs;][]{feinstein19}.

We ran \texttt{michael} on all stars in our samples from which flares were identified.
The estimated rotational periods were subsequently vetted by-eye  with the \texttt{michael}
diagnostic plots. This vetting was implemented to ensure that the measured $P_\textrm{rot}$
was not a harmonic of the true $P_\textrm{rot}$ or  from an occulting companion. This led
to robust measurements of rotation periods for \nprot\ stars in total. Additionally, we
identified 17 eclipsing binaries or potentially new planet candidates.

\section{Flare Rates of Young Stars}\label{sec:results}

In this section, we use the previously described methodology to estimate the flare
rates of the young stars in our sample. We implement three main steps in this analysis.
First, we perform the standard FFD fitting of a power-law to the distribution of
flare energies described in Section~\ref{subsec:energy_ffd}. Next we fit the relationship
between $R_0$ and flare rates in Section~\ref{subsec:rossby}. Finally we fit a
truncated power-law to the distribution of flare amplitudes in  Section~\ref{subsec3_truncated}
to identify correlations between $R_0$ and flare distributions.

The number of stars and flares in  each association varied greatly
(Figure~\ref{fig:flare_distribution}). This was primarily caused by the limited
number of stars that had been observed at 2-minute cadence in TESS.  We therefore
did not measure FFD properties as a function of association. Instead, we group
stars by $T_\textrm{eff}$ and average adopted association age.

We define the following spectral type bins by $T_\textrm{eff}$:
 \begin{itemize}
     \item M-stars below the fully convective boundary ($T_\textrm{eff} = 2300 - 3400$\,K),
     \item  early type M-stars ($T_\textrm{eff} = 3400 - 3850$\,K),
     \item  late K-stars ($T_\textrm{eff} = 3850 - 4440$\,K),
     \item early K-stars ($T_\textrm{eff} = 4440 - 5270$\,K),
     \item G-stars ($T_\textrm{eff} = 5270 - 5930$\,K).
 \end{itemize}
We did not include any stars hotter than $T_\textrm{eff} > 5930$\,K. These hot
stars generally exhibit light curves  dominated by noise in the TESS observations.
Additionally, we grouped stars in the following age space: $4-10$\,Myr (including
Upper Scorpius and TW Hydrae), $10-20$\,Myr, $20-40$\,Myr, $40-50$\,Myr, $70-80$\,Myr,
$120-150$\,Myr, and $150-300$\,Myr. We note that there is a gap in age from $50-70$\,Myr,
which could be expanded with the identification of more associations in this age range.

\subsection{Standard Power-Law Fits}\label{subsec:energy_ffd}

\begin{figure*}[bht!]
    \script{mcmc_results.py}
    \begin{centering}
        \includegraphics[width=\textwidth, trim={0cm 0 60cm 0}, clip]{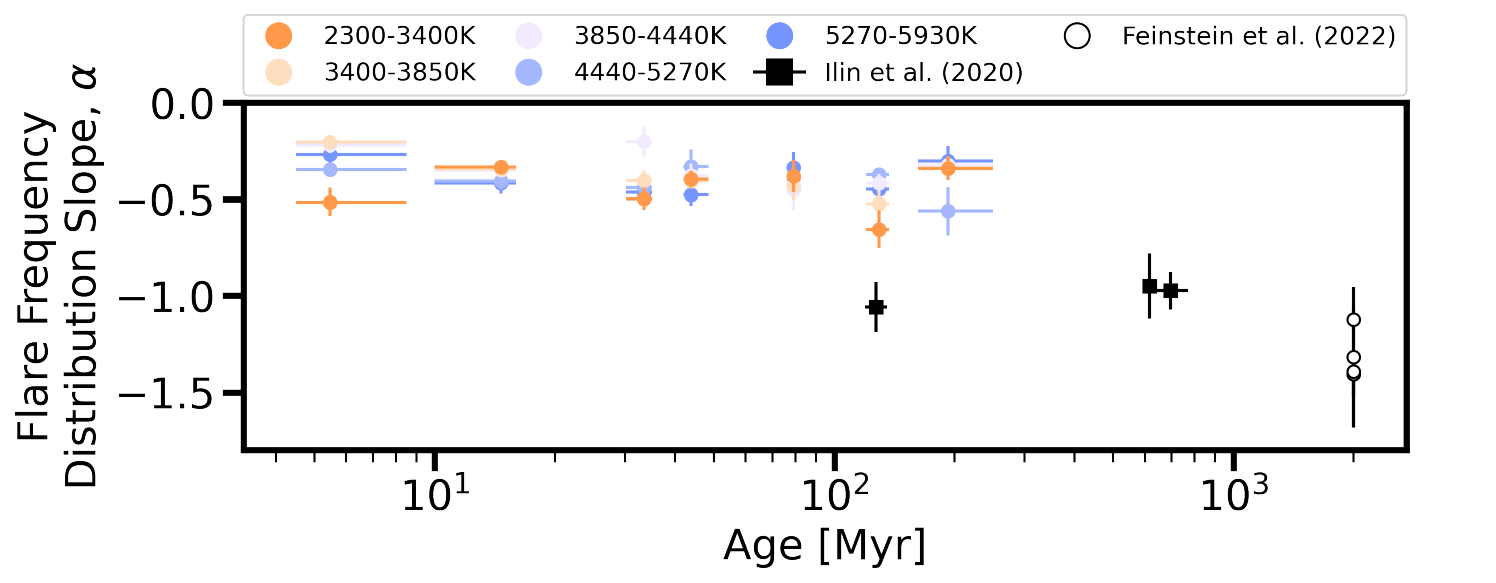}
        \caption{
            Measured flare-frequency distribution slopes, $\alpha$, as a function
            of stellar effective temperature, $T_\textrm{eff}$ and age. We measured
            these FFDs with respect to flare energy. We find the FFD slopes as a
            function of energy are consistent with $\alpha = -0.6 ~\textrm{to} -0.2$
            for stars $< 300$~Myr in TESS observations. A shallower FFD slope is
            indicative of more high-energy flares. We present all measured FFDs in
            Figure~\ref{appendix:supp_ffds}, and all measured slopes and errors in
            Table~\ref{tab:best_fits}. We find the shallowest slopes for stars
            $T_\textrm{eff} = 3400 - 4440$\,K, with a range from $\alpha = -0.44~ \textrm{to} -0.22$.
            We do not include the results for stars $T_\textrm{eff} = 3850 - 4440$\,K
            and $t_\textrm{age} = 20 - 40$\,Myr as this bin contained only six stars with
            detected flares. We present the average results of measured FFD slopes for
            the Hyades, Pleiades, and Praesepe clusters from \cite{ilin21} as black
            squares. We present the results of measured FFD slopes for all TESS primary
            mission targets in white circles \citep{Feinstein22} as ``field-age". We
            discuss what drives the difference between our sample and \cite{ilin21} in
            Section~\ref{subsec:energy_ffd}.
         }
        \label{fig:mcmc_results}
    \end{centering}
\end{figure*}

We fit the stars FFD slopes, approximated as a power-law, using the $T_\textrm{eff}$
and age bins described above. Flares were binned into 25 bins in log-space from
$10^{27} - 10^{35}$\,erg. The FFD has a notable turnover energy, $E_\textrm{turnover}$,
when our detection method is incomplete due the low-amplitude of those flares.
The FFD slope is often fit to flares with energies $E \geq E_\textrm{turnover}$.
We perform our fits following the same methodology. This turnover in the FFD
cannot be accurately modeled with a power-law. We present our FFDs as a function
of $T_\textrm{eff}$ and age in Figure~\ref{fig:simple_ffd_all}; points which were
fit to measure the FFD slope are presented in black, while the full FFD is presented
in gray.

We fit the FFD using the MCMC method implemented in \texttt{emcee} \citep{goodman10, emcee}.
Specifically, we fit for the slope, $\alpha$, y-intercept, $b$, and an additional
noise term, $f$. This noise term  accounts for an underestimation of the errors in
each bin. We initialized the MCMC fit with 300 walkers and ran our fit over 5000
steps.  We discarded the first 100 steps upon visual inspection, after which the
steps were fully burned-in. The measured FFD slopes, $\alpha$, are presented in
Figure~\ref{fig:mcmc_results}. We approximate the error on the slope as the lower
16\textsuperscript{th} and upper 84\textsuperscript{th} percentiles from the MCMC fit.

Overall, we measure a shallower FFD for stars of all masses at $t_\textrm{age} < 300$~Myr.
A shallower FFD indicates there are more high-energy flares. The measured FFD slopes
are shallower than those measured using smaller samples of young stars. \cite{jackman21}
fit the three FFDs for M3-M5, M0-M2, and K5-K8 stars younger than 40~Myr that had been
observed with Next Generation Transit Survey. They measured slopes of $0.94 \pm 0.04$,
$0.69 \pm 0.05$, $0.82 \pm 0.14$ per bin, which each contained $\leq 120$~flares.
It is also worth noting that the slopes measured here are based on up to an order of
magnitude more flares per fit. Our sample also has a longer temporal baseline,
allowing for the occurrence of more high-energy flares. Accumulating more high-energy
flares in the FFD, if such events occur, will result in a shallower FFD slopes. TESS
flare statistics for main sequence stars have indicated steeper FFDs \citep{feinstein22_criticality}.
Therefore, the FFDs presented here are not strictly impacted by longer observations,
but are the result of more high-energy flares on young stars.

There is a $1-3\sigma$ discrepancy between the FFD slope measured in \cite{ilin21}
and the work presented here at ages $\sim 120$\,Myr ($\alpha$ ranges from
$-1.32 \pm 0.19 ~\textrm{to} -0.91 \pm 0.18$, depending on the $T_\textrm{eff}$ bin).
\cite{ilin21} used the \textit{K2} 30-minute light curves for their analysis, compared
with our TESS 2-minute light curves. Additionally, our sample has $\sim 2 \times$
the number of stars and $\sim 7 \times$ the number of flares as in \cite{ilin21}.
First, we test if these discrepancies are driven by differences in sample binning.
We reevaluate the FFDs assuming the $T_\textrm{eff}$ bins presented in Table 3 of
\cite{ilin21}. We find $\alpha$ remains consistent with our presented values to
within $\sim 1\sigma$ of the nearest temperature bin. Second, we reevaluate the
FFDs assuming the total number of flares fit in \cite{ilin21}. We draw
$n_\textrm{fit, Ilin et al. (2021)}$ \citep[last column in Table 3 of ][]{ilin21}
flares from our sample 100 times without replacement, refit the FFD, and take
the average FFD slope from that sub-sample of flares to our results.  We find
$\alpha$ remains consistent with our presented values to $\sim 2\sigma$. We note
that in most cases $n_\textrm{fit, Ilin et al. (2021)} < n_\textrm{fit, this work}$.
Therefore, the increased disagreement in $\alpha$ could be due to smaller sample sizes.

Additionally, we test if the difference in $\alpha$ is driven by the cadence
differences between \textit{K2} (30-minute) and TESS (2-minute). We bin all of
the flares in our sample down to a 30-minute cadence, re-fit for flare $A$ and ED,
and recalculate the flare energy. We re-fit the FFD for this altered sample and
measure  FFD slopes consistent to $\sim 1 \sigma$ with those presented in this work.
Therefore, the discrepancy seen here is not driven by observational differences
between \textit{K2} and TESS.

Finally, we test if the differences are driven by the range of energies that are fit.
The CNN-based flare detection algorithm has previously been demonstrated to be
less-biased against lower energy flares. Like this work, \cite{ilin21} fit the FFD
across flare energies which are not affected by a reduced efficiency in the low-energy
flare detection, $E \geq E_\textrm{turnover}$. In energy space, these fits begin
between $E_f = 10^{32-33}$~erg, compared to our fits which begin between
$E_f = 10^{31-32}$~erg (Figure~\ref{appendix:supp_ffds}). When we limit our sample
to $E_f \geq 10^{32.5}$~erg, we find the FFD slopes become \textit{steeper} and
more consistent with the values of $\alpha$ presented in \cite{ilin21}. Therefore,
the difference in $\alpha$ between this work and \cite{ilin21} for stars with
$t_\textrm{age} = 120 - 150$~Myr is driven by the lower energy flares, with which
we are more complete to with TESS than \textit{K2}.

\subsection{Flare Rate Dependence on Rossby Number}\label{subsec:rossby}

\begin{figure*}[htb!]
    \script{prot_histograms.py}
    \begin{centering}
        \includegraphics[width=\textwidth]{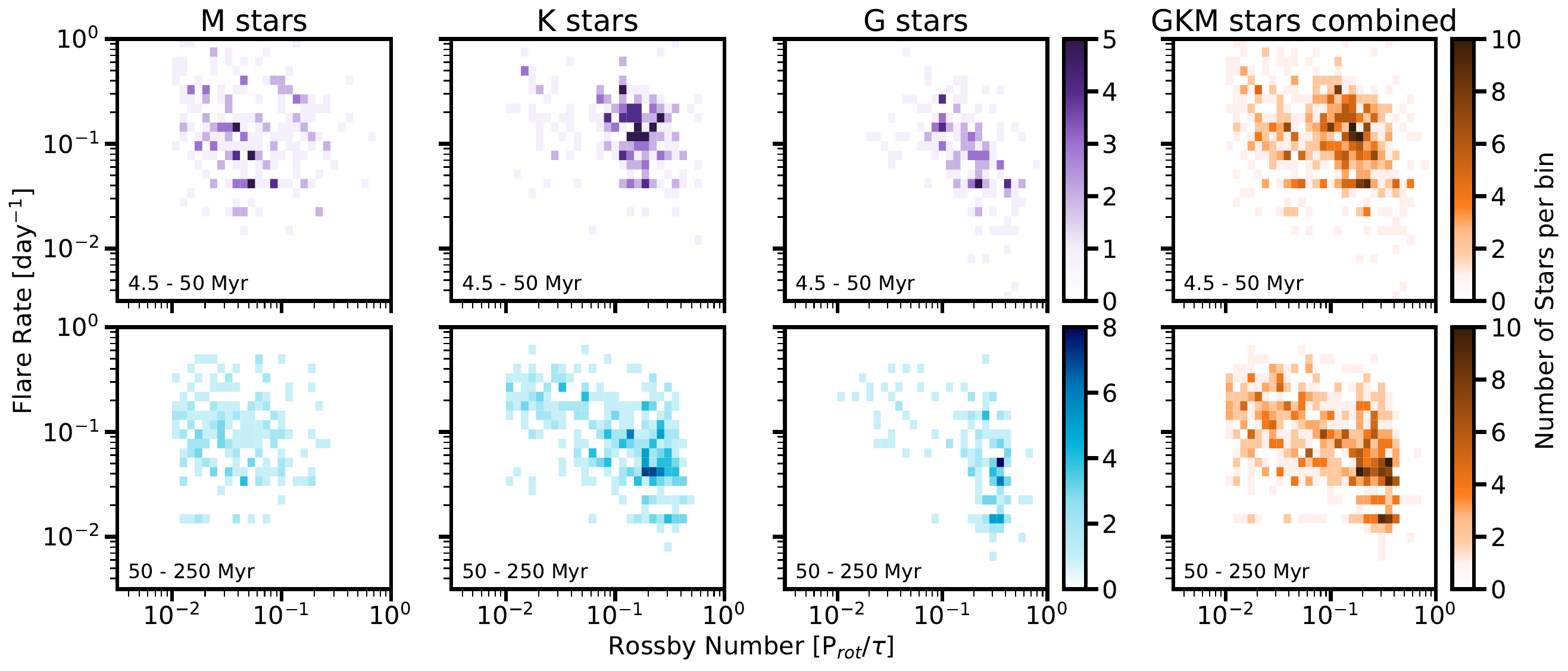}
        \caption{
            Comparison of Rossby Number, $R_0$ and flare rate for young GKM stars.
            For the younger sample (top row; $t_\textrm{age} = 4.5 - 50$~Myr), we
            find no correlations between flare rate and $R_0$. For the slightly
            older sample (bottom row; $t_\textrm{age} = 50 - 250$~Myr), we find
            no change in the average flare rate for M stars. For K and G stars,
            we start to see some evolution in this relationship. For K and G
            stars, we that as $R_0$ increases, the average flare rate decreases.
            This could indicate that as stars spin-down, their flare activity
            also begins to decline. We find that for the full GKM sample of
            stars (bottom row, rightmost column), the relationship between $R_0$
            and flare rate is best-fit by a broken power law, with a turnover
            at $R_0 = 0.136$. For the younger full sample (top row, rightmost column),
            we find this relationship is best-fit by a single power law. The
            histograms are colored by number of stars in each bin.
        }
        \label{fig:prot_histograms}
    \end{centering}
\end{figure*}

The Rossby number, $R_0$, is a parameter which incorporates  several properties which
are known to affect the stellar dynamo, such as the rotation period and stellar mass.
In the context of stellar dynamo the Rossby number indicates the dominance of the
convective verses rotational dynamo. It is defined as $R_0 = P_\textrm{rot}/\tau$,
where $\tau$ is the convective turnover time. We convert our measured rotation periods
to $R_0$,  approximating $\tau$ following the prescription in \cite{wright11}. We
equate the flare rates, $\mathcal{R}$, for individual stars as

\begin{equation}\label{eq:fr}
  \mathcal{R} = \frac{1}{t_\textrm{obs}} \left( \sum_{i=1}^{N} p_i \right)\,.
\end{equation}

In Equation (\ref{eq:fr}), $\mathcal{R}$ is the flare rate in units of day$^{-1}$,
$t_\textrm{obs}$ is the total amount of time a target was observed with TESS, and
$p_i$ is the probability that flare $i$ is a true flare as assigned by \texttt{stella}.
We compare the calculated $R_0$ to measured flare rates for all stars with measured
$P_\textrm{rot}$. The results are presented in Figure~\ref{fig:prot_histograms}.

From these results, we are able to evaluate the dependence of the flare rate on age,
spectral type,  and $R_0$. We split the sample between stars younger and older than
50~Myr. This age roughly correlates to the age at which GKM stars turn onto the main
sequence. The flare rate of stars younger than 50~Myr slightly decreases with
increasing $R_0$. However, there is a significant amount of scatter in this relationship.

As can be seen by comparing the upper and  lower panels in Figure ~\ref{fig:prot_histograms},
the flare rate dependence on Rossby number is most evident in older, more massive stars
between $50 - 250$~Myr. There is minimal evolution in both the average flare rate
and $R_0$ between the two samples for M stars. For K stars,  $R_0$ evolves more
dramatically during the first 250~Myr while the scatter in the flare rate decreases.
For G stars, the scatter in $R_0$ decreases and the average flare rate across the
sample decreases. We present a compiled histogram for all stars in our sample in the
right column of Figure~\ref{fig:prot_histograms}.

To better understand this trend for the GKM sample, we fit three functions to the
flare rate Rossby number parameter space: (i) a constant value, (ii) a single power
law, and (iii) a piece-wise function consisting of a constant value and a power law.
We computed the $\chi^2$ between each of these fits and the data. For (iii), we fit
for where the $R_0$ turnover should occur by computing the $\chi^2$ across a range
of $R_0 = [0.09, 0.18]$. We weighted the data points based on the density of points
in a given bin (Figure~\ref{fig:prot_histograms}). For  $4.5 - 50$~Myr old stars,
the distribution is best-fit with a single power law with slope $m = -0.102 \pm 0.018$
and y-intercept $b = -0.660 \pm 0.017$. For stars $50 - 250$~Myr, the distribution
is best-fit with a piece-wise function of the form:

\begin{equation}\label{eq:piecewiseft}
  \mathcal{R} =
  \Bigg \{
  \begin{array}{ll}
        C & R_0 \leq 0.136 \\
        10^b * R_0^m  & R_0 > 0.136 \\
  \end{array}
\end{equation}
In Equation (\ref{eq:piecewiseft}), $\mathcal{R}$ is the flare rate, $C = 0.269 \pm 0.007$,
$m = -0.612 \pm 0.039$, and $b = -1.113 \pm 0.035$. The location of the turnover is
consistent with what has been measured in other observations of magnetic saturation
for partially and fully convective stars \citep[e.g. $L_X/L_\textrm{bol}$; ][]{wright18}.

With a sample of 851,168 flares detected in both \textit{Kepler} 1- and 30-minute
cadence data, \cite{Davenport16} searched for a relationship between Rossby number
and relative flare luminosity. This work determined that the relationship could be
fit by a broken power law, with a break at $R_0 = 0.03$ and a slope of $m \sim -1$
dominating higher $R_0$. However, a single power law is slightly preferred over the
broken power law. While the metrics used between \cite{Davenport16} and this paper
are different, both suggest a saturation in flares for stars with low $R_0$. Additionally,
\cite{medina20} found a broken power-law relationship between $R_0$ and the flare
rate for flares with $E_f > 3.16 \times 10^{31}$\,erg, corresponding to the completeness
threshold of their sample. \cite{medina20} analyzed light curves of 419 low-mass
($M_\star = 0.1 - 0.3 M_\odot$) main-sequence stars in the Solar neighborhood
observed by TESS.  They found that the flare rate saturates at
$log(\mathcal{R}) = -1.30 \pm -0.08$ for $R_0 < 0.1$ and rapidly declines for $R_0 > 0.1$.
\cite{yang23} analysed 7,082 stars with measured $P_\textrm{rot}$ observed during
TESS sectors 1–30, and found a broken-power law described the relationship between
$\Delta$Flare$/L_\star$ for M dwarfs, with a break between $R_0 = 0.1 - 0.13$, which
is consistent with this work. While there are differences in sample selection between
our work and previous studies, all are consistent with flare rate saturation for
stars with low $R_0$ and a steep drop-off in flare rate for stars with higher $R_0 > 0.1$.

\subsection{Truncated Power-Law Fits}\label{subsec3_truncated}

We search for evidence of variations of the FFDs as a function of flare amplitude
versus Rossby number, $R_0$. This is motivated by the data  presented in \cite{seligman22}.
In that work, they  modeled flares driven by magnetic reconnection events driven by
rotational forces as well as convective dynamo.  \cite{seligman22} found that stars
with $R_0 < 0.13$ exhibited shallower FFD slopes than stars with $R_0 \geq 0.13$ to
several sigma significance. The differences in the FFDs were indicative of relatively
more high-energy flares to low-energy flares. This was interpreted as evidence of a
more dominant rotational dynamo compared to the convective dynamo, which preferentially
produced longer magnetic braids in the stellar coronae.

In this subsection, we test this theory with a much larger sample of stars (\nprot\
instead of $807$ stars). We fit the distributions separated by the fitted $R_0$
presented in Section~\ref{subsec:rossby}. In order to perform the fits, we follow
the prescription described in greater detail in \cite{seligman22}. Here, we provide
a brief summary of the prescription. Specifically, we fit a truncated power-law
distribution of the form,

\begin{equation}\label{eq:dpda}
  dp/dA \propto A^{-\alpha_T} e^{-A/A_*}\,.
\end{equation}

\begin{figure}[ht!]
    \script{truncated.py}
    \begin{centering}
        \includegraphics[width=\linewidth]{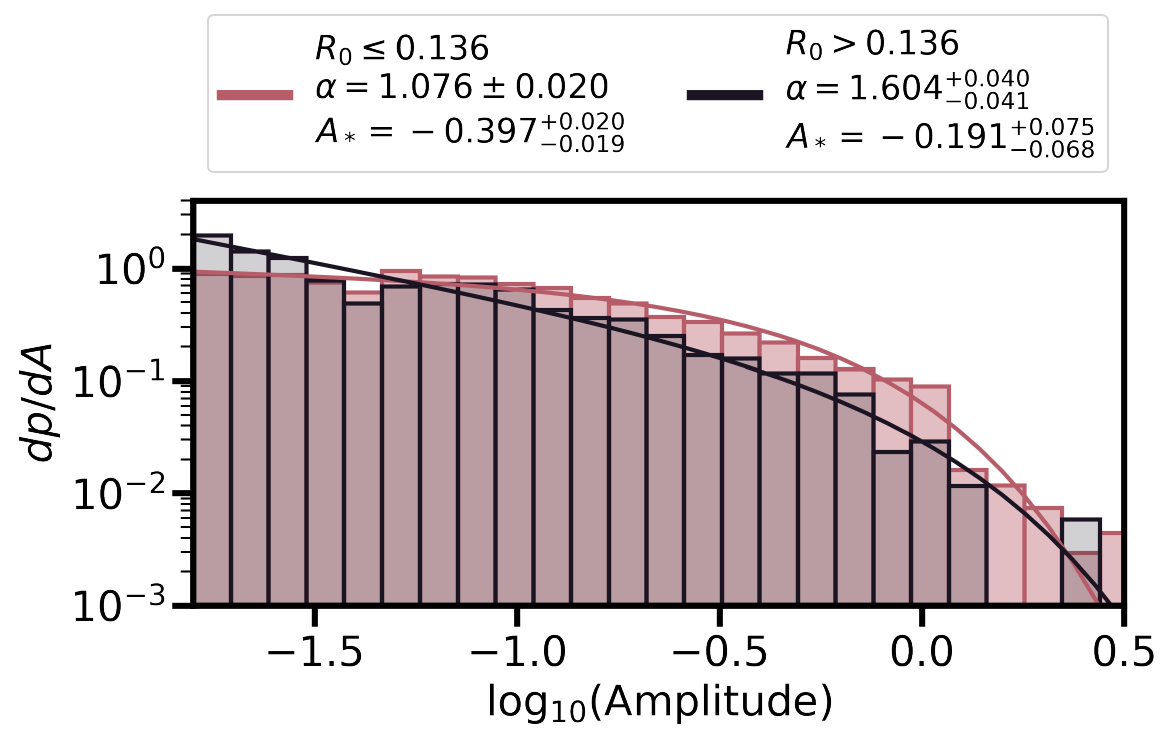}
        \caption{
            Flare frequency distributions, as a function of flare amplitude,
            for stars with $R_0 \leq 0.12$ (red) and stars with $R_0 > 0.12$
            (black). We present the best-fit model, and the best-fit values of
            the model slope and normalization factor in the legends. Our sample
            of $R_0 \leq 0.12$ includes 13,132 flares from 800 stars; our sample
            of $R_0 > 0.12$ includes 5,603 flares from 747 stars. We find the stars
            with smaller Rossby numbers have shallower slopes, consistent with more,
            high-energy flares and a more dominant rotational dynamo \citep{seligman22}.
        }
        \label{fig:truncated}
    \end{centering}
\end{figure}

In Equation (\ref{eq:dpda}),  $A$ is the amplitude of the flare, $A_*$ is a flare
amplitude cutoff parameter and $\alpha_T$ is the slope, rather than $\alpha$. We fit
the slopes using the MCMC method implemented in \texttt{emcee} \citep{goodman10, emcee}.
We used the log-likelihood function in \cite{seligman22} and  fit for $A_*$ and $\alpha_T$.
We initialized the MCMC fit with 200 walkers and evaluated the fit over 5000 steps.
The first 1000 steps were discarded upon visual inspection. The results are presented
in Figure~\ref{fig:truncated}.

We find that stars with $R_0 \leq 0.136$ have a best-fit slope of
$\alpha_T = 1.076 \pm 0.020$, while stars with larger $R_0$ have a best-fit slope
of $\alpha_T = 1.604 \pm 0.040$. This result agrees with the results presented in
\cite{seligman22}. This is not the first instance we have seen a correlation in FFDs
as a function of $R_0$. \cite{candelaresi14} noted that faster rotating stars have
higher superflare rates derived from \textit{Kepler} data. The spectroscopic survey
presented in \cite{notsu19} found that the maximum flare energy decreased as
$P_\textrm{rot}$, and consequently $R_0$, increased. \cite{mondrik19} identified
flares across 2,226 stars observed with MEarth and found an increase in flare rate
between stars with low $R_0 < 0.04$ and stars with intermediate $R_0 = (0.04, 0.44)$,
and a subsequent decrease in flare rate for stars with high $R_0 > 0.44$. While our
bins of stars by $R_0$ are not as fine as those presented in \cite{mondrik19}, the
correlation in $\Delta R_0$ remains consistent.

\section{Evidence for Stellar Cycles from Variable Flare Activity}\label{sec:cycles}

The Sun undergoes an 11-year solar cycle during  which it oscillates between high and
low magnetic activity. This magnetic cycle manifests  in a variety of observables. One
of the primary indicators of the solar cycle is a stark change in the flare rate; this
rate can vary by more than an order of magnitude between Solar maxima and minima
\citep{webb94}. Moreover, energetics of the flares that are produced on the solar
surface vary dramatically over the course of the solar cycle \citep{Bai1987,Bai2003}.
Direct and indirect evidence  of the solar cycle have also been observed in radio flux,
total solar irradiance, the magnitude and geometry of the magnetic field, coronal
mass ejections affiliated with flares, geomagnetic activity, cosmic ray fluxes and
radioisotopes in ice cores and tree rings. For a recent review, see \citet{Hathaway2015}.

While other stars should also experience activity cycles like the Sun, they are
more difficult to measure. Constraints on stellar cycles have
predominantly relied on long baseline variations in stellar photometry, typically
observed at cadences which cannot resolve stellar flares \citep[see recent reviews
by ][]{jeffers23, isik23}. However,  tracing stellar cycles via stellar flares may
be more reliable, as flares are a direct consequence of magnetic activity. \cite{scoggins19}
explored measuring the stellar cycle length of KIC 8507979, a star in the \textit{Kepler}
field which was observed for 18 90-day quarters. \cite{scoggins19} found the flare rate
decreased over each quarter, which could be fit by
$L_\textrm{fl}/L_\textrm{Kp} = (-9.96 \pm 3.94) \times 10^{-2} \times t_\textrm{yr} + (2.43 \pm 0.11)$,
where $t_\textrm{yr}$ is the time in years, and $L_\textrm{fl}/L_\textrm{Kp}$ is
a parameterization of the \textit{Kepler} flare rate \citep{lurie15}.

\subsection{Flare Observables from the Sun}

Here, we explore observables from solar flares between 2002 and 2018 in the Reuven
Ramaty High Energy Solar Spectroscopic Imager \citep[RHESSI;][]{lin02} flare
catalog.\footnote{\url{https://hesperia.gsfc.nasa.gov/rhessi3/data-access/rhessi-data/flare-list/index.html}}
The RHESSI mission observes the Sun across a wide range of X-ray energies,
from 3 keV-17 MeV, with high temporal and energy resolutions as well as with high signal
sensitivity. Such a wide energy range enables the ability to explore both thermal and
non-thermal emission observed during flare events. Over 16 years of operations, over
$100,000$ solar flares have been observed and characterized in RHESSI data. These
observations covered the second half of solar Cycle 23 and the beginning of Solar
Cycle 24. We use the publicly available RHESSI flare catalog to determine  metrics
that would be most useful when searching for evidence of stellar cycles. One potential
issue with this analysis is that the HXR observations from RHESSI and white-light
observations from TESS may not be directly correlated. However, there is a close spatial
and temporal correspondence between HXR and white-light flares on the Sun
\citep{fletcher07, krucker11, kleint16}. In particular, \cite{namekata17}
demonstrated the difference spatially resolved flares as observed with RHESSI HXR and the
Solar Dynamics Observatory (SDO)/Helioseismic and Magnetic Imager (HMI) white-light. In
some examples, the HMI white-light emission is seen to last longer than the HXR,
indicating the white-light emission is related to non-thermal electrons.

\begin{sidewaysfigure*}[ht!]
    \vspace{10cm}
    \script{stellar_cycles.py}
    \begin{centering}
        \includegraphics[width=\textwidth]{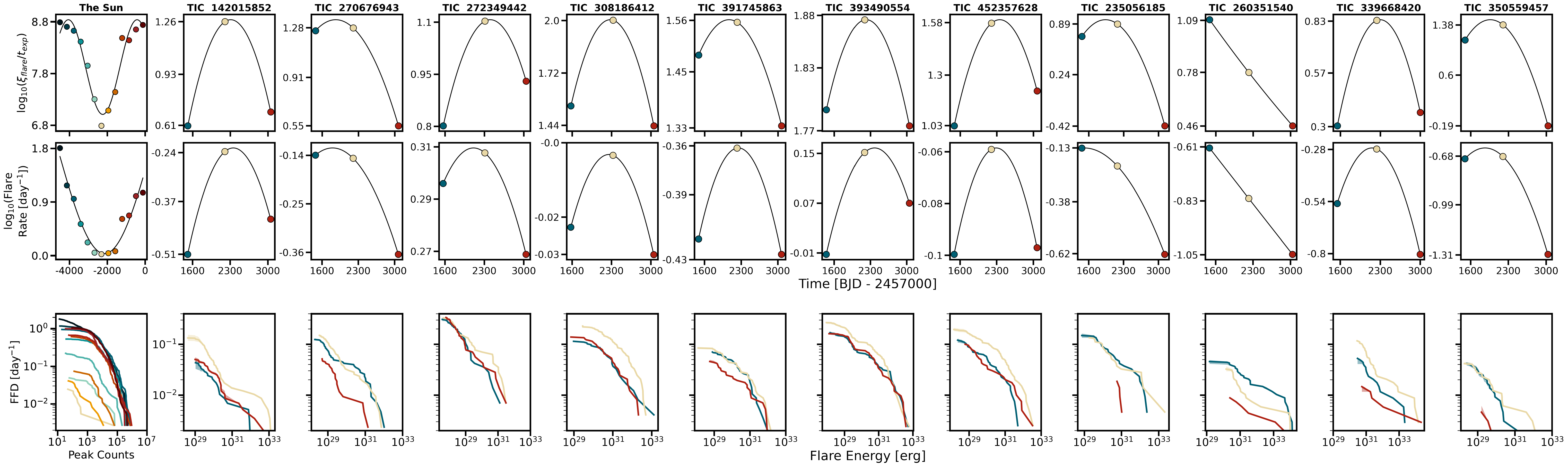}
        \caption{
            Comparison of three flare-driven metrics in searching for evidence of stellar cycles. Each column contains information for the Sun (left), followed by TIC 142015852,
            269797536, 270676943, and 278825715. The top row shows trends in the fractional luminosity emitted in flares, $L_{fl}/L_\textrm{TESS} \equiv \xi_\textrm{flare}/t_\textrm{exp}$
            \citep{lurie15}. The second and third rows show the calculated flare rate and cumulative flare frequency distribution (FFD) respectively. For the Sun, we used the RHESSI X-ray
            flare catalog and binned the flares per year from February 2002 - February 2014. The Sun shows clear trends in $\xi_\textrm{flare}/t_\textrm{exp}$ and flare rate over the 12-year
            solar cycle. The cumulative FFD shows that (i) the overall number of flares decreases, which can be seen in the flare rate as well and (ii)  the high-energy flare end of the
            distribution decreases. A quadratic polynomial is plotted to help guide the eye. It is important to note that these fits do not provide constraints on the length of potential
            stellar cycles in these stars due to the limited baseline obtained with TESS thus far. We present the measured flare rates and fractional luminosity emitted for each star in Table~\ref{tab:cycles}
        }
        \label{fig:stellar_cycles}
    \end{centering}
\end{sidewaysfigure*}

However, other examples of simultaneous HXR and white-light flare
observations in \citep{namekata17} show similar flare durations. Therefore, we take
the HXR solar flares with energies between $25 - 50$~keV as likely representative
of solar white-light flares, and are a comparable sample to our TESS sample. First,
the total number of flares detected in the $25 - 50$~keV bandpass varied by 650~flares
from solar maximum in 2002 to solar minimum in 2008. This variation correlates to
a change in flare rate from $1.81$ to $0.02$ flares day$^{-1}$ on average. Second,
the flare frequency distribution changes between solar minimum and maximum because
the frequency of the highest-energy flares decreases.  Finally, the total flare
luminosity relative to the total luminosity correlates with the stellar cycle.
This is defined by \cite{lurie15} as

\begin{equation}
    \frac{L_\textrm{flare}}{L_\star} \equiv \frac{\xi_\textrm{flare}}{t_\textrm{exp}}
\end{equation}

where $L_\textrm{flare}$ is the total luminosity emitted by flares, $L_\star$ is
the luminosity of the star, $\xi_\textrm{flare}$ is sum of the equivalent duration
of all flares observed, and $t_\textrm{exp}$ is the exposure time of the observations.
These results for the Sun are presented in the left-most column of Figure~\ref{fig:stellar_cycles}.

\subsection{Quantifying Flare Activity over 5 Years of TESS Observations}

The TESS Extended Missions have provided a five-year baseline that we can use to
search for evidence of stellar cycles. This is a comparable baseline to \textit{Kepler},
although with sparser sampling. We search for evidence of stellar cycles  within
our sample of fast rotators. Our sample contains 108 stars which have been observed
for $t_\textrm{obs} \geq 200$\,days across five years. We search for evidence of
changes in the stellar magnetic activity by looking for correlations in (i)
variations in the total luminosity emitted in flares  (ii) the total flare rate
per year and (iii) annual changes in the FFD.  We group our observations by the
year in which the observations were taken, even for the cases where the star was
not continuously observed throughout that year (e.g. Sectors 1-26 are Year 1,
Sectors 27-55 are Year 2, and Sectors 56-67 are Year 3.

We searched our 108 candidates by eye for correlations in the aforementioned criteria.
The correlations are categorized into the following four options: (i) \textbf{Positive}:
$E_\textrm{f, max}$ and the total number of flares has increased over five years; (ii)
\textbf{Negative}:  $E_\textrm{f, max}$ and the total number of flares has decreased
over five years; (iii) \textbf{Trough}: $E_\textrm{f, max}$ and the total number of
flares was greater in the first and third years observed with a minima in the second
year; (iv) \textbf{Peak}:  $E_\textrm{f, max}$ and the total number of flares was
fewer in the first and third years observed with a maxima in the second year.

From these three observables, we identified eleven stars which all display evidence
of scenario Peak, with the exception of one star displaying evidence of scenario
Negative, which are presented in Figure~\ref{fig:stellar_cycles}. These stars are TIC
142015852, 235056185, 260351540, 270676943, 272349442, 308186412, 339668420, 350559457,
391745863, 393490554, and 452357628.  Our sample includes seven early type M-stars,
two late K-stars, one early K-star, and one G-star. TIC 272349442 is a candidate member
of TW Hydrae ($t_\textrm{age} = 10 \pm 3$\,Myr), TIC 142015852, 270676943, 308186412,
339668420, 350559457, and 452357628 are candidate members of Carina
($t_\textrm{age} = 45 \pm 9$\,Myr), TIC 235056185 and  260351540 are candidate members
of IC 2602 ($t_\textrm{age} = 52.5 \pm 2.95$\,Myr), and TIC 391745863 and 393490554
are candidate members of the AB Doradus Moving Group ($t_\textrm{age} = 133 \pm 12.5$\,Myr).

We report the measured $\xi_\textrm{flare}/t_\textrm{exp}$ and flare rates across
all three years in presented in Figure~\ref{fig:stellar_cycles} in Table~\ref{tab:cycles}.
Across our sample, we find the largest change in flare parameters in TIC 235056185,
which has a $\Delta log_{10}(\xi_\textrm{flare} / t_\textrm{exp}) = 1.14$ and
$\Delta \mathcal{R} = 0.5$. Additionally, we see some targets (e.g. TIC 350559457)
become very flare quiet, going from $\mathcal{R} = 0.2$~day$^{-1}$ to $\mathcal{R} = 0.05$~day$^{-1}$,
which is obvious when visually inspecting the light curve. For stars showing the
Peak scenario, we find that the year before and after the peak do not necessarily
return to the same value of $\xi_\textrm{flare} / t_\textrm{exp}$ and $\mathcal{R}$,
although it does for some stars.

\subsection{Flare Recovery per TESS Year}\label{subsec:inj-rec}

The majority of stars identified with evidence of stellar cycles exhibit the Peak
scenario. To determine if this is astrophysical or driven by some unknown instrumental
systematic, we perform an injection-recovery test on this sub-sample of stars.
Injection-recovery tests are used to address biases in our detection method.
Injection-recovery tests are typically not recommended for machine-learning
detection techniques \citep{feinstein20}. However, we perform them because these
are some of the first results of searching for evidence of stellar cycles via
flare activity.

We inject a total of 50 flares into each sector of data for our eleven stars.
We draw the amplitude of our flares from a Gaussian distribution centered at a
3\% increase in flux, with a standard deviation of 1\%. We do not inject flares
with amplitudes below 0.5\%. We used the Llamaradas Estelares flare model. We
fit a line between flare amplitude and FWHM from our new catalog to extract an
appropriate FHWM for any given amplitude. We added additional noise to the FHWM,
as the relationship between amplitude and FWHM exhibits scatter similar to
distribution of flare energy and ED shown in Figure~\ref{fig:flare_distribution}.
Once the flares were injected, we followed the steps outlined in
Sections~\ref{subsec2:Flareidentification} - \ref{subsec2:qualitychecks}
to identify these injected flares.

We considered a flare recovered if the $t_\textrm{peak}$ is within 15~minutes of
the injected $t_\textrm{peak}$. We are able to recover 93\% of flares with
$A \geq 3$\% and 80\% of flares with $A < 3$\%. This is consistent with previous
results using this flare identification method \citep{feinstein20}. We note that
there is a steep drop-off in flare recovery rate at $T_\textrm{mag} > 13$, with
average recovery rates dropping from  90\% to 70\%.  The variation in recovery rates
between years observed by TESS ranges from $1 - 8$\% across all eleven targets.
There is no correlation between recovery rate,  $\xi_\textrm{flare}/t_\textrm{exp}$
and flare rate, with the exception of TIC 142015852 and 339668420. We find an 8\%
increase between Years 1 and 2, and a 3\% decrease between Years 2 and 3 in the
recovery rate for TIC 142015852. Additionally, we find an 1\% increase between
Years 1 and 2, and a 2\% decrease between Years 2 and 3 in the recovery rate for
TIC 339668420. An 8\% change in the recovered flare rate correlates to
$\Delta (\xi_\textrm{flare} / t_\textrm{exp}) = 0.09$, assuming ED~$= 1$\,minute,
and a $\Delta \mathcal{R} = 0.03$. These estimates are smaller than the annual
variations seen in flare statistics for these two stars, leading us to conclude
that these variations are astrophysical.

\subsection{Observing Peaks in Stellar Flare Activity}

In the Section~\ref{subsec:inj-rec} we demonstrated that 10 out of 108 stars with
$> 200$~days of observations exhibited Peak behavior in their variability patterns
and found no bias in flare detection as a function of year. This is approximately
what we would expect to find for an unbiased sample. Our analysis requires a large
number of bright flares to be observable to characterize the stellar cycle. Therefore,
our sample of objects should be biased towards stars that are undergoing a local
maximum, rather than a local miinimum, in their activity cycle.

It is not clear what the length of the activity cycles are for stars other than
the Sun. For this order of magnitude estimate we therefore assume that the activity
cycles of all stars in our sample are similar to that of the Sun. This approximation
is valid   if the length of stellar cycles on average is not more than an order
of magnitude different than that of the sun. By observing $N\sim$108 stars for a
$\tau\sim$5 year baseline, assuming that each has a $P\sim10$ year activity cycle,
we would expect that during that time period $\tau /P N\sim50$ stars should
experience a peak in their stellar cycle during the observations. Moreover, our
analysis will only be sensitive to activity cycles if there is not only a peak,
but also lower activity in the year before and after the peak. Therefore, the
number of stars that we would expect to identify that have a peak in year 2 should
be $\tau /P N /3\sim16$. Therefore we would expect to identify $\sim16$ stars
exhibiting a peak in their activity cycle during these observations, which is
similar to what we have found.

However, it is likely that there is a variety of lengths of stellar cycles in our
population and the exact correspondence between the number found and the number
estimated is simply a manifestation of the crude order of magnitude approximation
that we employed rather than an exact correspondence. Nevertheless, this represents
tentative detections of peaks in stellar activity cycles and that on average the
length of these cycles is not an order of magnitude different than that of the Sun.

The results presented here are consistent with those presented in
previous studies of magnetic activity cycles in young stars. \cite{olah16} analyzed
29 stars from the Mount Wilson survey which had $\sim 36$ years of \ion{Ca}{2}
H \& K monitoring. Roughly 16 stars in this sample have constrained ages of $< 1$~Gyr
via gyrochronology. This sample of stars have $P_\textrm{rot} = 18.1 \pm 12.2$~days
and magnetic activity cycles of $P_\textrm{cyc} = 7.6 \pm 4.9$~years. There is
significantly more dispersion in the young stellar cycle lengths than for the older
stars. \cite{olah16} concluded that young stars show more complex interannual variations
in magnetic activity. While stars age, magnetic braking will increase the rotation
of the star. Once this process has occurred, the magnetic activity cycle becomes more
well-behaved, as seen on the Sun.

We apply the relationship between $log(1/P_\textrm{rot})$ and
$log(P_\textrm{cyc}/P_\textrm{rot})$ described in \cite{olah16} to estimate the
activity cycles in our sample. 10 of our 11 candidates have measured
$P_\textrm{rot} = 0.41 - 5.22$~days. Using the best-fit slope of $0.76 \pm 0.15$,
we estimate the $P_\textrm{cyc} = 0.81 - 1.49$~years. Strong evidence of shorter
activity cycles than $\sim 3$~years may be missed by the sparse annual sampling
from TESS. Redesigning the next TESS extended mission to stare at the northern and
southern ecliptic poles continuously for two years, instead of alternating poles,
may help resolve these shorter activity cycles, if present.

\subsection{Validating Stellar Activity Cycles}

The stars presented in this work present examples that stellar activity cycles could be identified by characterizing stellar flares and measuring flare rates. In addition to flare rates, it would
be beneficial to conduct detailed spectroscopic follow-up of these candidates. For example, detailed monitoring of \ion{Ca}{2} H \& K lines for these targets could reveal if their activity cycles
are similar to that seen from the flare rates. Unfortunately, our candidates are located too far south to have been included in the Mount Wilson HK project, which aimed to understand stellar
chromospheric activity on a variety of timescales \citep{Wilson68}. Monitoring of the overall XUV luminosities of these targets could provide additional insights into the variability of the targets.
Additionally, more continuous time-series photometry of these targets would provide more stringent constraints on the flare variability for these targets. While long time-series photometry,
such as that achieved with TESS and \textit{Kepler} are ideal, more sparse but consistent photometric monitoring for stellar flares may provide useful additional supplemental data to TESS observations.
This is especially true for the times when TESS is observing in the opposite ecliptic hemisphere.

The TESS Full-Frame Images (FFIs) allow for the creation of light curves for objects in $\sim 95\%$ of the sky \citep[e.g.][]{feinstein19}. During its primary mission (July 2018 - July 2020),
the TESS FFIs were taken every 30-minutes. During its first extended mission (July 2020 - September 2022), the FFI cadence was decreased to 10-minutes. Now, during its second extended mission
(September 2022 - September 2025), the FFI cadence was again decreased to 3-minutes. At 3-minutes, we can more accurately identify and  resolve the structure of stellar flares \citep{howard22}.
A detailed search of flare variability for stars in the TESS FFIs may provide a more statistical view of how flare rates change on long timescales, and if they can yield insights into stellar
activity cycles.

\section{Discussion}\label{sec:discuss}

\subsection{Correlations with Far- and Near-Ultraviolet Flux}

The stellar X-ray coronal emission is known to be a tracer of overall magnetic activity. Therefore,  it should theoretically depend on the dynamo of the star, and on observable parameters that
are related to the dynamo such as the rotational period. Empirical evidence indicates that there are distinct saturated and non-saturated regimes for coronal X-ray emission from main sequence
stars. Specifically, stellar X-ray luminosity surveys have revealed that the saturation limit depends on the stellar rotation period. Namely, there is no evolution in $L_X/L_\textrm{bol}$ for
stars with $P_\textrm{rot} < 10$\,days \citep{Pizzolato03}. The rotation period is a good predictor of the X-ray luminosity for stars with longer $P_\textrm{rot}$.

The Far- and Near-Ultraviolet (FUV/NUV) emission is another tracer of magnetic activity. Younger, more active stars display excess luminosity in both of these wavelengths \citep{Shkolnik2013}.
We use archival observations from the \textit{Galaxy Evolution Explorer} \citep[\textit{GALEX};][]{martin05} to search for trends in FUV/NUV saturation and flare rate saturation, similar to the
established X-ray trends. \textit{GALEX} provides broad FUV photometry from $1350-1750 \angstrom$ and NUV photometry from $1750-2750 \angstrom$. We crossmatch our target stars with the
\textit{GALEX} catalog following the sample selection methods outline in \citep{schneider18}. Specifically, we search using a $10''$ radius around the coordinates of each target in our sample.
We include targets with no bad photometric flags (e.g. \texttt{fuv\_artifact} or \texttt{nuv\_artifact == 0}) as defined
in the catalog. It is recommended by the \textit{GALEX} documentation to exclude any objects with these flags associated. Additionally, we exclude targets with measured magnitudes brighter than
15, which marks the saturation limit for both the FUV and NUV photometers \citep{morrissey07}.Based on these thresholds, we find that 462 stars in our sample have NUV photometry and 139 stars have
FUV photometry.

\begin{figure}[ht!]
    \script{galex.py}
    \begin{centering}
        \includegraphics[width=\linewidth]{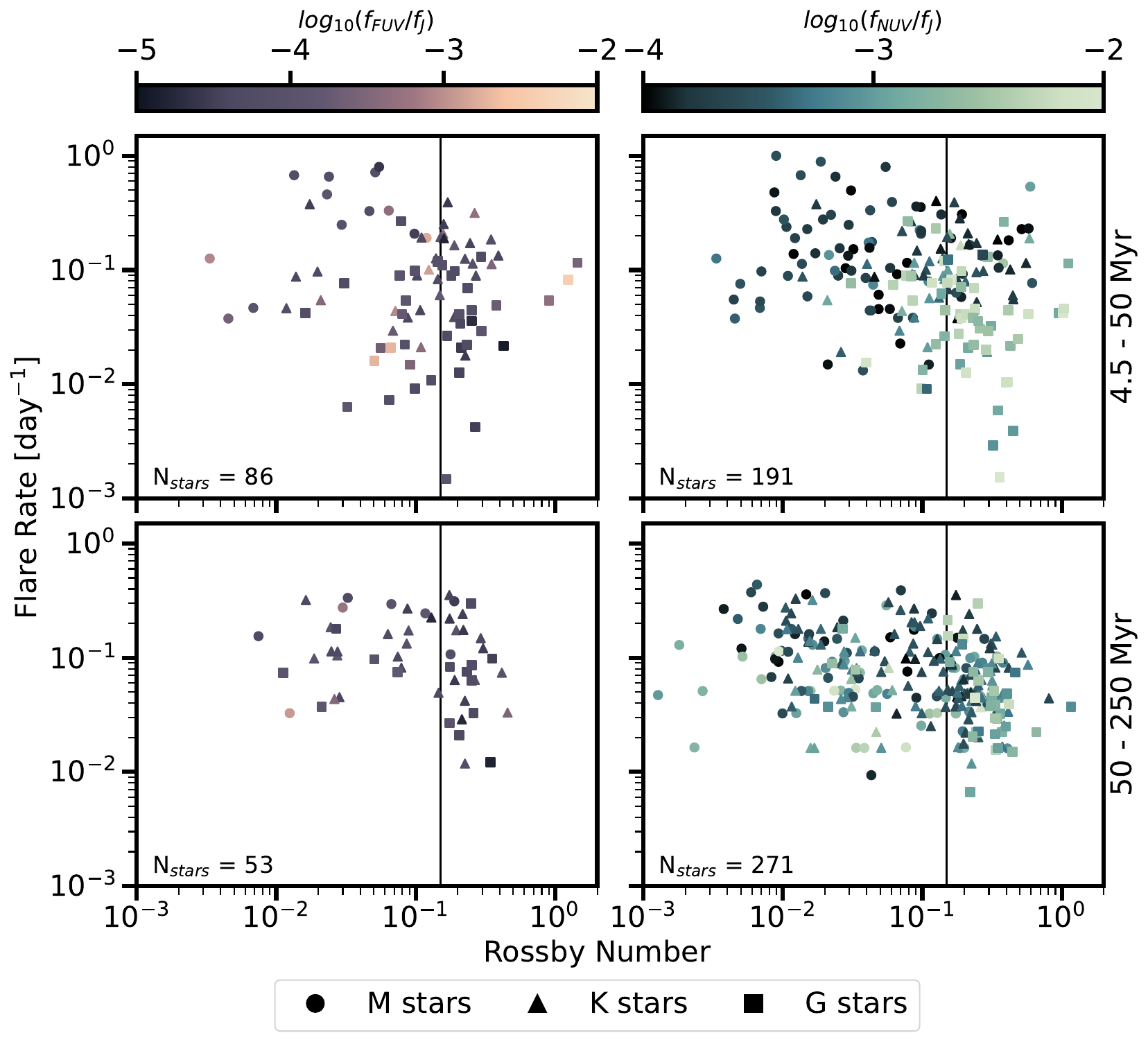}
        \caption{
            Calculated FUV (left) and NUV (right) GALEX flux for stars in our
            sample normalized  by the stellar J-band flux.
            There is no obvious correlation between the fractional NUV/FUV flux and the measured
            flare rate or Rossby number.
            The top row shows GKM stars $<50$~Myr; the bottom row
            shows GKM stars $\geq 50$~Myr. M stars are shown as circles, K stars as
            triangles, and G stars as squares. We note that $f_{NUV}$ traces the
            photosphere of G, K, and massive M stars, and therefore may not be the best comparison
            bandpass when looking for trends in magnetic activity.
        }
        \label{fig:galex}
    \end{centering}
\end{figure}

We investigate whether the saturation of flare rate and FUV/NUV emission are correlated with the derived $R_0$ in each star. We present our results comparing the FUV and NUV flux and the measured
flare rate and Rossby number in Figure~\ref{fig:galex}. We present the measured FUV/NUV flux normalized by the J-band flux, which acts as an activity indicator. In theory, the bolometric luminosities
should be a better normalization factor than the J-band flux. However, the majority of stars in our sample do not have bolometric luminosity  measurements. Therefore, we keep the normalization to
$f_J$ while assessing FUV/NUV correlations for the larger statistical sample.

While there is tentative evidence of trends between the flare rate and the Rossby number (see Section~\ref{subsec:rossby}), there is inconclusive evidence that the normalized FUV and NUV flux
follow this trend. We note that  the number of stars in our sample which have \textit{GALEX} FUV measurements is small ($N_\textrm{stars} = 139$) and therefore might not be representative of the
broader population. It is worth noting that the NUV flux traces the photosphere for many of these stars, as opposed to the X-rays which trace coronal emission. Additionally, the FUV still has
contributions from the photosphere for G stars. Therefore, it is possible that the lack of a correlation between the regimes identified for X-ray emission is due to the fact that the UV is tracing
different stellar atmospheric regions that are not associated with the magnetic activity. 
not hold for FUV and NUV flux.

The FUV and NUV flux from \textit{GALEX} is a superposition of many different emission lines. These lines  trace various regions of the stellar atmosphere ranging from the corona to the
photosphere depending on their formation temperatures. It is possible that the blending of lines produces the lack of correlation between FUV/NUV flux, $R_0$, and flare rate. Spectroscopic
observations of targets with $R_0$ which span the transition out of the saturated regime may reveal a stronger relationship between these parameters. \cite{pineda21} reported evidence of broken
power-law relationship between $R_0$ and FUV emission lines for $\sim 20$ stars observed with the Space Telescope Imaging Spectrograph on the \textit{Hubble} Space Telescope. Depending on the
emission line analysed, \cite{pineda21} found a saturated regime for $R_0 < 0.18 - 0.24$, with a steep drop-off for higher $R_0$. Additionally, \cite{loyd21} evaluated the relationship between
FUV emission lines and $R_0$ for 12  Tucana-Horologium ($t_\textrm{age} = 40$~Myr), 9 Hyades  ($t_\textrm{age} = 650$~Myr), and 7 field-aged ($t_\textrm{age} = 2-10$~Gyr) M stars. They found a
saturated $log_{10}(R_0) = -0.876_{-0.061}^{+0.037}$, which is consistent with the X-ray flux and $R_0$ relationship.

The ROentgen Survey with an Imaging Telescope Array (eROSITA) instrument \citep{predehl07, predehl21} on the Russian Spectrum-RG mission is an all-sky X-ray survey from $\sim 0.2-8$~keV. The
synergies between eROSITA and TESS are already being explored. \cite{magaudda22}, used the measured $L_X$ from eROSITA and $P_\textrm{rot}/R_0$ from TESS for 704 M dwarfs to reconfirm the known
X-ray activity relationship. It is possible that these combined data may show a  clearer relationship between $R_0$ and flare rate in the X-ray than shown here  in the FUV/NUV (Figure~\ref{fig:galex}).

\subsection{Flare Rates of Young Planet Host Stars Verses Comparison Sample}

The all-sky observing strategy of TESS has revealed a new population of young transiting exoplanets.  Characterization of the  environment of these planets is crucial to understanding their
subsequent evolution. Specifically, the stellar environment can impact how these young planets evolve into their mature counterparts.

It is unclear whether stellar flares are beneficial or detrimental to the habitability of exoplanets. It is possible that stellar flares can trigger the development of prebiotic chemistry
\citep{Rugheimer2015,airapetian16,Ranjan2017,Rimmer2018}. On the other hand, stellar flares and affiliated coronal mass ejections can permanently alter atmospheric compositions \citep{chen21}.
This alteration may increase the amount of atmospheric mass stripped during the early stages of planet evolution \citep{feinstein20}. Therefore,  understanding the environment of young
transiting exoplanets can  provide insight into their evolution.

To this end we compare  measured flare rates of young, planet hosting stars to a statistical sample of stars with similar ages and $T_\textrm{eff}$ that do not have confirmed short-period planets.
Specifically, we measured the flare rates of planet hosting stars with ages $< 300$\,Myr, comparable to the ages of our primary sample. We followed the methods outlined in Section~\ref{sec:methods}
to detect and vet flares for the planet-hosting stars.

\begin{figure}[ht!]
    \script{yp_rates.py}
    \begin{centering}
        \includegraphics[width=\linewidth]{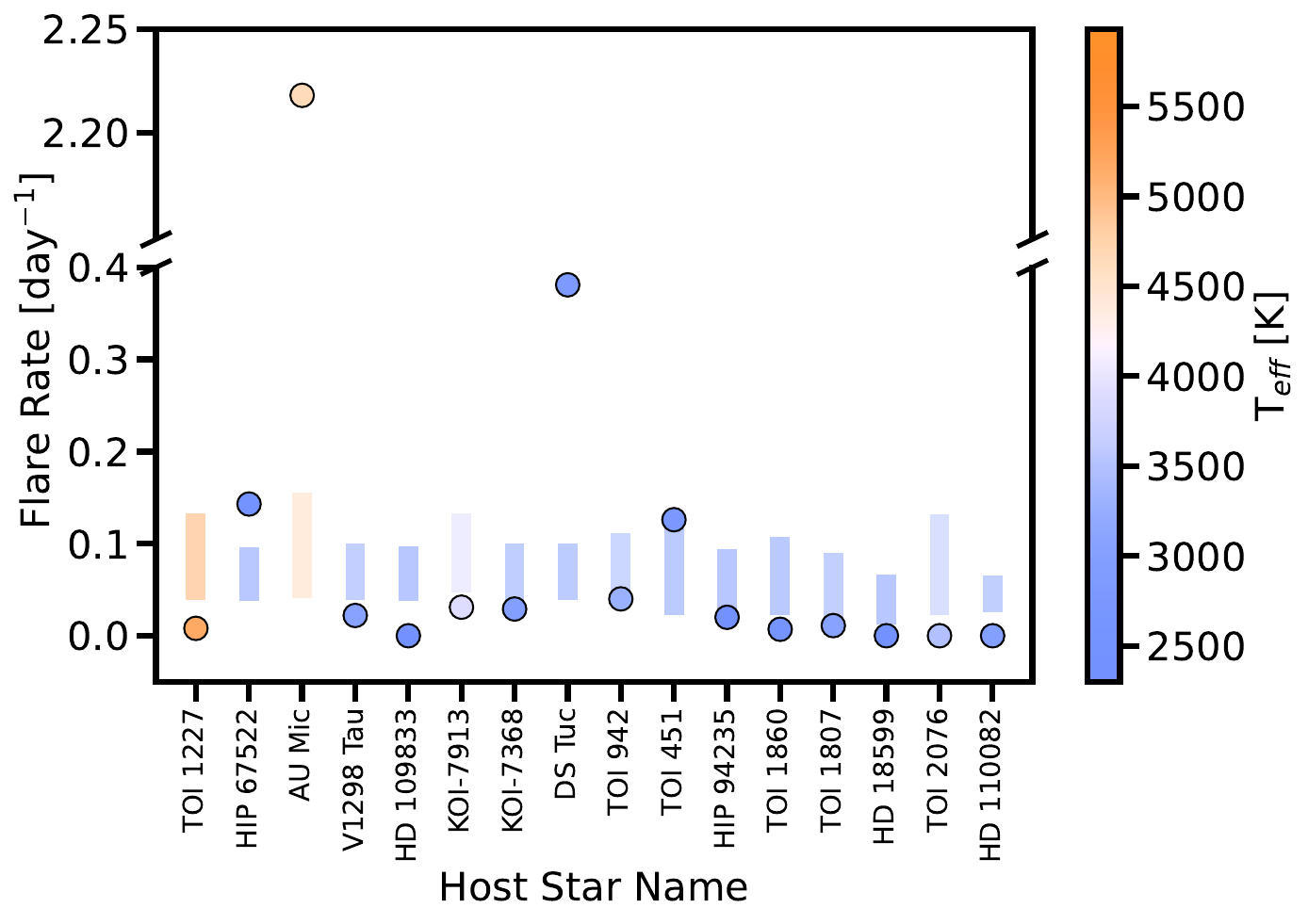}
        \caption{
            Comparison of flare rates from young planet host stars with respect
            to a comparable sample with respect to age [Myr] and $T_\textrm{eff}$
            [K]. Circles represent the flare rate of the host star (name along the x-axis);
            vertical bars represent the lower 16\textsuperscript{th} and upper
            84\textsuperscript{th} percentiles for the comparable sample. The
            majority of young planet host stars have comparable flare rates to
            the lower end of the comparison sample. A handful of hosts have more
            flares, including HIP 67522, AU Mic, DS Tuc A, and TOI 451. We find no
            correlation with spectral type or age which may indicate why these host
            stars are relatively flare quiet. The measured flare rates are presented in
            Table~\ref{tab:yp_rates}.
        }
        \label{fig:yp_rates}
    \end{centering}
\end{figure}

For our comparison sample, we considered stars with ages $\pm 30$\,Myr of the planet hosting star and $T_\textrm{eff} \pm 1000$\,K. For each of the stars in the control samples, we calculate the
flare rate following Equation~\ref{eq:fr}. We present the flare rates of planet-hosting stars and a comparable sample of stars in Figure~\ref{fig:yp_rates} and report the measured rates in
Table~\ref{tab:yp_rates}. For the control samples, we report the median flare rate, and the lower 16\textsuperscript{th} and upper 84\textsuperscript{th} percentiles.

Flare rates are slightly diminished for the majority of planet hosting stars compared to the control sample. This could be interpreted as evidence that the presence of planets inhibits magnetic
reconnection events and activity, or that the presence of flares biases transit-detection algorithms. However, for all cases where the flare rates are diminished in the presence of a short-period
planet, the difference between the flare rates are within 1$-\sigma$ and not statistically significant.

There are a handful of cases where stars hosting short-period planet exhibit drastically higher flare rates. The most dramatic case is for the 23 Myr AU Mic \citep{jeffries05,malo14,Mamajek14},
where the flare rate is more than an order of magnitude higher than in the control sample.  AU Mic and its transiting planets are generally considered as a benchmark laboratory for understanding
the impact of stellar activity on young exoplanet atmospheres. The system hosts an  extended debris disk \citep{kalas04, liu04, metchev05} along with \textit{two} short-period transiting planets
\citep{plavchan20, martioli21, gilbert22}. It is worth noting that this flare rate is consistent with that measured by \citet{gilbert22, feinstein22_aumic}, and is not attributed to star-planet
interactions \citep{ilin22}. In addition to this, HIP 67522, DS Tuc A, and TOI 451 all exhibit higher flare rates than the comparison sample.

It is unclear what differentiates the planet hosting systems with higher flare rates from the control sample. \cite{france18} and \cite{behr23} found that planet hosting stars are less active
than an equivalent control sample of field age stars in the UV. In Figure~\ref{fig:yp_rates} the color of the points corresponds to the effective temperature of the star. It appears that the
flare rates of the planet hosting stars compared to the control sample is randomly distributed with respect to the effective temperature of the star. It is possible that there is a small age
effect: the systems with higher flare rates are some of the youngest planet hosting stars: HIP 67522 is $17 \pm 2$ Myr, AU Mic is $22 \pm 3$ Myr, DS Tuc is $45 \pm 4$  Myr and TOI 451 is $120 \pm 10$ Myr.
There is a slight preference for the younger systems to exhibit elevated flare rates, but this is marginal evidence at best. Future observations of these systems with elevated flare rates may
reveal what causes this feature.

\begin{deluxetable}{l r r r r}[!ht]
\tabletypesize{\footnotesize}
\tablecaption{Young Planet Host Flare Rates\label{tab:yp_rates}}
\tablehead{
\colhead{Host Name} & \colhead{Age} & \colhead{Flare Rate} & \colhead{Comp. Sample} & \colhead{$N$} \\
 & \colhead{[Myr]} & \colhead{[day$^{-1}$]} & \colhead{Flare Rate [day$^{-1}$]}
}
\startdata
TOI 1227 &  $11 \pm 2$ &  0.008  & $0.065_{-0.042}^{+0.074}$ & 168 \\
\textbf{HIP 67522} & $17 \pm 2$ &  0.169  & $0.081_{-0.043}^{+0.099}$ & 368 \\
\textbf{AU Mic} & $22 \pm 3$ &  2.218  & $0.104_{-0.066}^{+0.188}$ & 590 \\
V1298 Tau & $23 \pm 4$ &  0.022  & $0.116_{-0.07}^{+0.107}$ & 300 \\
HD 109833 & $27 \pm 3$ & 0.000  & $0.077_{-0.039}^{+0.097}$ & 415 \\
KOI-7913 & $36 \pm 10$ &  0.031  & $0.152_{-0.094}^{+0.165}$ & 312 \\
KOI-7368 & $36 \pm 10$ &  0.029 & $0.116_{-0.071}^{+0.107}$ & 285 \\
\textbf{DS Tuc} & $45 \pm 4$ &  0.420  & $0.104_{-0.061}^{+0.104}$ & 281 \\
TOI 942 & $50_{-20}^{+30} $ &  0.040  & $0.092_{-0.049}^{+0.146}$ & 120 \\
\textbf{TOI 451} & $120 \pm 10$ &  0.128  & $0.059_{-0.03}^{+0.136}$ & 204 \\
HIP 94235 & $133_{-20}^{+15}$ & 0.020 & $0.048_{-0.028}^{+0.094}$ & 165 \\
TOI 1860 & $133 \pm 26$ &  0.008  & $0.049_{-0.026}^{+0.108}$ & 207 \\
TOI 1807 & $180 ± 40$ &  0.013  & $0.06_{-0.038}^{+0.11}$ & 53 \\
HD 18599 & $200_{-70}^{+200} $ & 0.000 & $0.033_{-0.02}^{+0.066}$ & 22 \\
TOI 2076 & $204 \pm 50$ & 0.000 & $0.083_{-0.041}^{+0.122}$ & 55 \\
HD 110082 & $250_{-70}^{+50} $ & 0.000 & $0.088_{-0.045}^{+0.045}$ & 2 \\
\enddata
\tablecomments{$N$ is the total number of stars used to calculate the comparable sample flare rate. We highlight host stars with  flare rates higher than those in the comparison sample.}
\end{deluxetable}

\section{Conclusions}\label{sec:conclusions}

In this work, we present the first measured flare rates for stars $< 300$\,Myr using TESS 2-minute cadence observations. We identified \nflares\ flares from \nflarestars\ stars (Figures~\ref{fig:sample} and \ref{fig:flare_distribution}). The results of our work are summarized as follows:

\begin{enumerate}
  \item We measured the flare-frequency distribution (FFD) slope, $\alpha$, for samples of flares binned by age and $T_\textrm{eff}$. We find $\alpha$ saturates at $\alpha = -0.6 ~ \textrm{to} -0.2$ for stars younger than 300\,Myr and declines after that age (Figure~\ref{fig:mcmc_results}). This is the first evidence that flare rates saturate across spectral types, as do  other tracers of stellar magnetic activity,.

  \item We measured rotation periods for \nprot\ stars and find that the relationship between flare rate and Rossby number, $R_0$, is best described as a piece-wise function with a turnover at $R_0 = 0.136$ for stars $t_\textrm{age} > 50$\,Myr (Figure~\ref{fig:prot_histograms}). Additionally, we find that stars with $R_0 \leq 0.136$ have a shallower FFD than stars with $R_0 > 0.136$, which is evidence of a more dominant rotational dynamo compared to the convective dynamo (Figure~\ref{fig:truncated}); this is consistent with results presented in \cite{seligman22}.

  \item We searched for evidence of far- and near-Ultraviolet (FUV/NUV) flux saturation as a function of $R_0$, similar to what is seen in the X-ray, by cross matching our sample with the \textit{GALEX} catalogs. We find no correlation between the FUV and NUV flux with flare rate (Figure~\ref{fig:galex}). The NUV (and for the G-type stars, the FUV) flux traces the photosphere for many of the stars in our sample, unlike the X-ray which traces the corona.  Spectrally-resolved NUV and FUV observations where the chromospheric emission lines can be  isolated may be a more promising means of investigating this connection. Future synergies between eROSITA and TESS may reveal such relationships as well.

  \item We compared the flare rates of planet-hosting young stars with a comparable sample with $t_\text{age} = t_\textrm{age, host} \pm 30$\,Myr and $T_\textrm{eff} = T_\textrm{eff, host} \pm 1000$\,K. We find that the majority of planet-hosting stars are flare inactive relative to a larger population of similar stars (although not to a statistically significant level), with the exception of HIP~67522, AU~Mic, DS~Tuc, and TOI~451 (Figure~\ref{fig:yp_rates}).

  \item We searched for evidence of long-term stellar cycles by evaluating changes in flare rates and FFDs  over five years of TESS observations. We identified ten candidates which show potential evidence of a local maxima in their stellar cycle, and one candidate which shows a decline in flare activity (Figure~\ref{fig:stellar_cycles}). We determine that these maxima are not due to flare-detection biases via injection recovery tests. While we are unable to obtain stellar cycle timescales from three data points, these results highlight the insights flares can bring to understanding stellar dynamos for targets with more TESS observations (e.g. stars in the continuous viewing zones) and the use of future TESS extended missions.

\end{enumerate}

\vspace{3mm}

\section{acknowledgements}

We thank David Wilson for thoughtful conversations. We thank the anonymous reviewer for their insightful report,
which improved the clarity and quality of this manuscript. This work made use of the open-source package,
\textcolor{red}{\textit{showyourwork!}} \citep{luger2021}, which promotes reproducible publications. ADF acknowledges funding from
NASA through the NASA Hubble Fellowship grant HST-HF2-51530.001-A awarded by STScI. DZS is supported by an NSF Astronomy and Astrophysics
Postdoctoral Fellowship under award AST-2202135. This research award is partially funded by a generous gift of Charles Simonyi to the NSF
Division of Astronomical Sciences. The award is made in recognition of significant contributions to Rubin Observatory’s Legacy Survey of Space and Time.

\appendix
\restartappendixnumbering

\section{Supplemental Material}\label{appendix:supp_ffds}

\begin{figure*}[ht!]
    \begin{centering}
        \includegraphics[width=\textwidth]{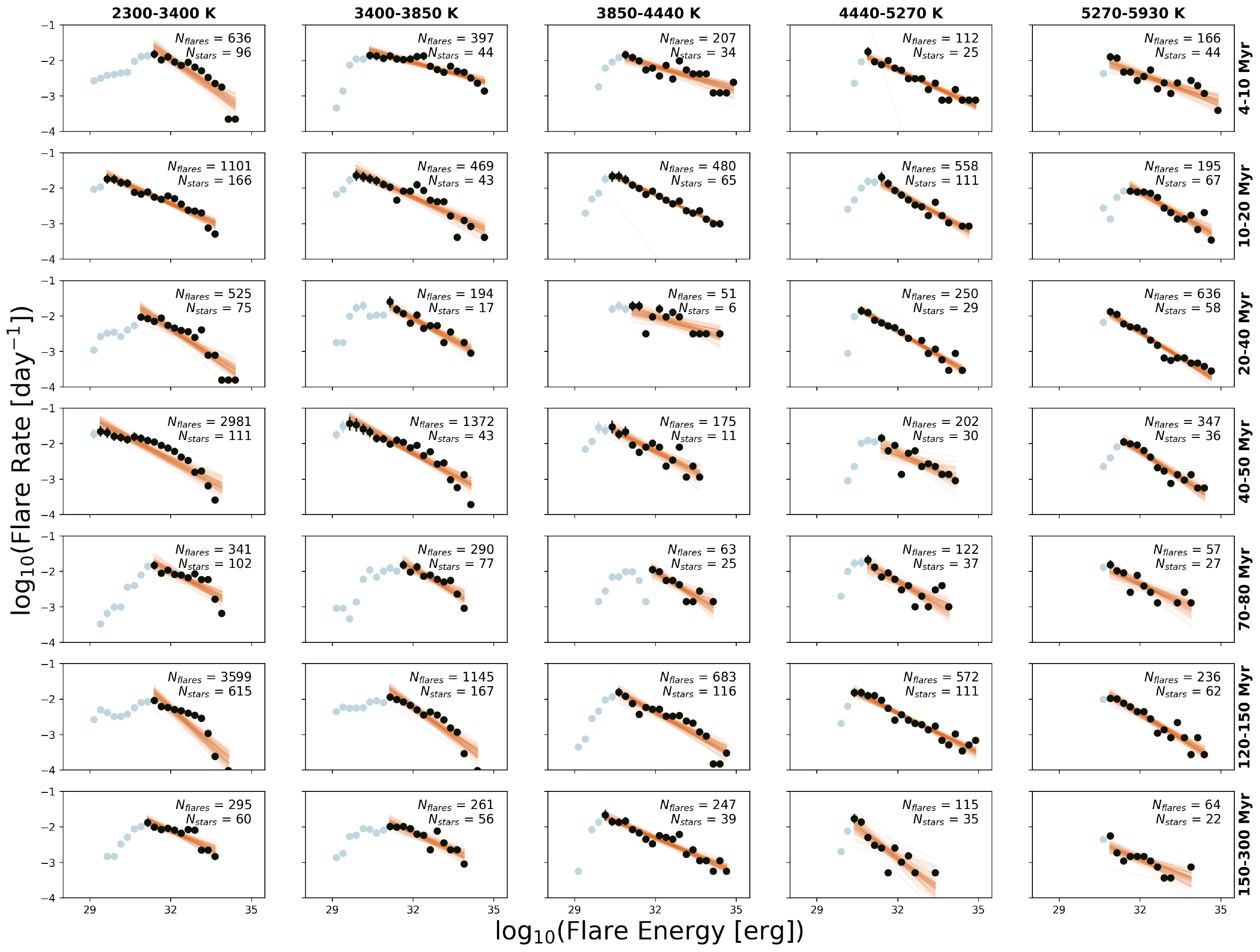}
        \caption{
            Flare frequency distributions (FFDs) for subgroups of stars, clustered
            by age and effective temperature, $T_\textrm{eff}$. Flares were sorted
            into 25 bins in log-space from $10^{27} - 10^{35}$\,erg. We fit the FFD
            from the turn-over in the binned flares, likely a result of  low-energy
            flares  missed by the flare-detection algorithm. The bins used to fit
            the FFD are shown in black, while all bins are shown in gray. We ran an MCMC
            fit to these distributions with a simple power law; 100 random samples from these
            fits are over-plotted in orange. We fit distributions with $> 3$ bins.
            The best-fit slopes from these fits are presented in Figure~\ref{fig:mcmc_results}.
        }
        \label{fig:simple_ffd_all}
    \end{centering}
\end{figure*}

In this appendix we present all of our best-fit slopes and intercepts for various flare-frequency distributions (FFDs) fit throughout this work. The fits outlined in Sections~\ref{subsec:energy_ffd} are included. We present the FFDs as a function of energy in Figure~\ref{fig:simple_ffd_all}.  100 random draws from our MCMC fitting to these distributions are shown as orange lines. We fit each distribution with $E > 10^{29}$\,erg, which roughly represents the turnover in each distribution. We do not fit the slope for $T_\textrm{eff} = [3850 - 4440]$~K at 20 -- 40~Myr due to our limited sample size (6 stars in total).

\vspace{3mm}

\begin{deluxetable*}{l | r | r r r r r r }[!ht]
\tabletypesize{\footnotesize}
\tablecaption{Best-fit slope and normalization parameters for stars of different temperatures and ages\label{tab:best_fits}}
\tablehead{
\colhead{$T_\textrm{eff}$ [K]} & \colhead{Age [Myr]} & \colhead{$\alpha_\textrm{E}$} & \colhead{$\sigma_{\alpha_\textrm{E}}$} & \colhead{$\beta_\textrm{E}$} & \colhead{$\sigma_{\beta_\textrm{E}}$} &  \colhead{$n_\textrm{E, fit}$} & \colhead{$N_\textrm{E}$}
}
\startdata
2300 -- 3400 & 4 -- 10  &  -0.516  &  0.074  &  14.514  &  2.409  &  731  &  96 \\
& 10 -- 20  &  -0.336  &  0.031  &  8.296  &  0.973  &  1183  &  166 \\
& 20 -- 40  &  -0.499  &  0.058  &  13.564  &  1.858  &  526  &  75 \\
& 40 -- 50  &  -0.397  &  0.045  &  10.248  &  1.400  &  3358  &  111 \\
& 70 -- 80  &  -0.383  &  0.083  &  10.244  &  2.685  &  341  &  102 \\
& 120 -- 150  &  -0.657  &  0.097  &  18.757  &  3.161  &  3797  &  615 \\
& 150 -- 300  &  -0.341  &  0.063  &  8.799  &  2.029  &  295  &  60 \\
\hline
3400 -- 3850  & 4 -- 10  &  -0.206  &  0.025  &  4.536  &  0.821  &  397  &  44 \\
& 10 -- 20  &  -0.345  &  0.038  &  8.768  &  1.199  &  474  &  43 \\
& 20 -- 40  &  -0.405  &  0.054  &  10.890  &  1.741  &  194  &  17 \\
& 40 -- 50  &  -0.408  &  0.037  &  10.770  &  1.182  &  1420  &  43 \\
& 70 -- 80  &  -0.432  &  0.075  &  11.893  &  2.471  &  291  &  77 \\
& 120 -- 150  &  -0.525  &  0.062  &  14.557  &  1.998  &  1185  &  167 \\
& 150 -- 300  &  -0.339  &  0.058  &  8.683  &  1.904  &  262  &  56 \\
\hline
3850 -- 4440  & 4 -- 10  &  -0.221  &  0.040  &  4.884  &  1.314  &  207  &  34 \\
& 10 -- 20  &  -0.350  &  0.016  &  8.998  &  0.545  &  480  &  65 \\
& 20 -- 40  &  -0.202  &  0.077  &  4.425  &  2.492  &  51  &  6 \\
& 40 -- 50  &  -0.379  &  0.062  &  9.873  &  1.982  &  175  &  11 \\
& 70 -- 80  &  -0.455  &  0.102  &  12.498  &  3.347  &  63  &  25 \\
& 120 -- 150  &  -0.419  &  0.046  &  11.010  &  1.494  &  684  &  116 \\
& 150 -- 300  &  -0.323  &  0.029  &  8.026  &  0.929  &  247  &  39 \\
\hline
4440 -- 5270  & 4 -- 10  &  -0.348  &  0.031  &  8.850  &  1.014  &  112  &  25 \\
& 10 -- 20  &  -0.408  &  0.043  &  10.914  &  1.417  &  558  &  111 \\
& 20 -- 40  &  -0.440  &  0.033  &  11.635  &  1.075  &  250  &  29 \\
& 40 -- 50  &  -0.331  &  0.089  &  8.325  &  2.893  &  202  &  30 \\
& 70 -- 80  &  -0.390  &  0.085  &  10.152  &  2.749  &  122  &  37 \\
& 120 -- 150  &  -0.375  &  0.029  &  9.593  &  0.930  &  572  &  111 \\
& 150 -- 300  &  -0.560  &  0.126  &  15.056  &  3.951  &  115  &  35 \\
\hline
5270 -- 5930  &  4 -- 10  &  -0.271  &  0.049  &  6.280  &  1.586  &  166  &  44 \\
& 10 -- 20  &  -0.419  &  0.055  &  11.266  &  1.820  &  195  &  67 \\
& 20 -- 40  &  -0.465  &  0.035  &  12.377  &  1.149  &  636  &  58 \\
& 40 -- 50  &  -0.479  &  0.054  &  13.071  &  1.780  &  347  &  36 \\
& 70 -- 80  &  -0.337  &  0.083  &  8.416  &  2.656  &  57  &  27 \\
& 120 -- 150  &  -0.450  &  0.045  &  11.977  &  1.460  &  236  &  62 \\
& 150 -- 300  &  -0.305  &  0.081  &  6.826  &  2.598  &  64  &  22 \\
\enddata
\tablecomments{$n_\textrm{E, fit}$ is the number of flares fit per bin; $N_E$ is the number of stars in each bin.}
\end{deluxetable*}

\begin{deluxetable*}{l | r r r | r r r}[!ht]
\tabletypesize{\footnotesize}
\tablecaption{Evidence of stellar cycles from variable flare properties \label{tab:cycles}}
\tablehead{
\colhead{TIC} & \colhead{} & \colhead{log$_{10}(\xi_\textrm{flare}/t_\textrm{exp})$} & \colhead{}  & \colhead{} & \colhead{$\mathcal{R}$ [d$^{-1}$]} \\
\hline
\colhead{} & \colhead{Year 1} & \colhead{Year 2} & \colhead{Year 3}  & \colhead{Year 1} & \colhead{Year 2} & \colhead{Year 3}
}
\startdata
142015852 & 0.61 & 1.26 & 0.69 & 0.31 & 0.58 & 0.38 \\
270676943 & 1.26 & 1.28 & 0.55 & 0.72 & 0.71 & 0.43 \\
272349442 & 0.80 & 1.10 & 0.93 & 1.98 & 2.03 & 1.86 \\
308186412 & 1.54 & 2.00 & 1.43 & 0.95 & 0.99 & 0.93 \\
391745863 & 1.49 & 1.56 & 1.33 & 0.38 & 0.44 & 0.37 \\
393490554 & 1.79 & 1.88 & 1.77 & 0.97 & 1.42 & 1.17 \\
452357628 & 1.03 & 1.58 & 1.21 & 0.80 & 0.87 & 0.80 \\
235056185 & 0.73 & 0.89 & -0.41 & 0.74 & 0.61 & 0.24 \\
260351540 & 1.09 & 0.78 & 0.46 & 0.24 & 0.15 & 0.09 \\
339668420 & 0.30 & 0.83 & 0.37 & 0.28 & 0.53 & 0.16 \\
350559457 & 1.15 & 1.38 & -0.19 & 0.20 & 0.21 & 0.05\\
\enddata
\end{deluxetable*}

\software{%
    numpy \citep{numpy},
    matplotlib \citep{matplotlib},
    scipy \citep{jones01},
    lightkurve \footnote{\url{https://doi.org/10.5281/zenodo.2557026}}, banyan-$\Sigma$ \citep{gagne18},
    astropy \citep{astropy:2013, astropy18}, \texttt{stella} \citep{stella_joss, feinstein20}, tensorflow \citep{Abadi2016}, astroquery \citep{astroquery19}
    }

\facility{TESS}

\bibliography{bib}

\end{document}